\documentclass[fleqn,usenatbib]{mnras}

\usepackage{newtxtext,newtxmath}

\usepackage[T1]{fontenc}

\DeclareRobustCommand{\VAN}[3]{#2}
\let\VANthebibliography\thebibliography
\def\thebibliography{\DeclareRobustCommand{\VAN}[3]{##3}\VANthebibliography}

\usepackage{graphicx}	
\usepackage{amsmath}	
\usepackage{ulem}

\title[X-ray polarimetry of compact-object obscurers]{X-ray polarization properties of partially ionized equatorial obscurers around accreting compact objects}

\author[J. Podgorn\'y et al.]{
J. Podgorn\'y$^{1,2,3}$\thanks{E-mail: jakub.podgorny@asu.cas.cz},
F. Marin$^{1}$ 
and M. Dov{\v{c}}iak$^{2}$ 
\\
$^{1}$Universit\'e de Strasbourg, CNRS, Observatoire Astronomique de Strasbourg, UMR 7550, F-67000 Strasbourg, France\\
$^{2}$Astronomical Institute, Academy of Sciences of the Czech Republic, Bo{\v{c}}n\'i II, CZ-14131 Prague, Czech Republic\\
$^{3}$Astronomical Institute, Charles University, V Hole{\v{s}}ovi{\v{c}}k\'ach 2, CZ-18000 Prague, Czech Republic\\
}

\date{Accepted XXX. Received YYY; in original form ZZZ}

\pubyear{2023}

\begin{document}
\label{firstpage}
\pagerange{\pageref{firstpage}--\pageref{lastpage}}
\maketitle

\begin{abstract}
We present the expected X-ray polarization signal resulting from distant reprocessing material around black holes. Using a central isotropic power-law emission at the center of the simulated model, we add distant equatorial and axially symmetric media that are covering the central accreting sources. We include partial ionization and partial transparency effects, and the impact of various polarization and steepness of the primary radiation spectrum. The results are obtained with the Monte Carlo code {\tt STOKES} that considers both line and continuum processes and computes the effects of scattering and absorption inside static homogenous wedge-shaped and elliptical toroidal structures, varying in relative size, composition and distance to the source. We provide first order estimates for parsec-scale reprocessing in Compton-thin and Compton-thick active galactic nuclei, as well as winds around accreting stellar-mass compact objects, for observer's inclinations above and below the grazing angle. The resulting reprocessed polarization can reach tens of \% with either parallel or perpendicular orientation with respect to the axis of symmetry, depending on subtle details of the geometry, density and ionization structure. We also show how principal parameters constrained from X-ray spectroscopy or polarimetry in other wavelengths can lift the shown degeneracies in X-ray polarization. We provide an application example of the broad modelling discussion by revisiting the recent IXPE 2--8 keV X-ray polarimetric observation of the accreting stellar-mass black hole in Cygnus X-3 from the perspective of partial transparency and ionization of the obscuring outflows.
\end{abstract}




\section{Introduction}\label{introduction}

Since the establishment of the unification scenario of active galactic nuclei (AGN) \citep{Antonucci1993,Urry1995}, the physical and -- in particular -- geometrical description of AGNs has reached much greater detail. The central supermassive black hole is believed to be surrounded by an accretion disc at sub-parsec scales with a dusty obscuring material reaching a few parsecs further out in the equatorial plane. Polarimetry not only allowed the discovery of the global AGN morphology back in the 60's and 80's \citep{Dibai1966, Walker1966, Angel1976, Antonucci1982, Antonucci1985}, but also serves as a powerful microscope into these extragalactic objects. That is because polarization -- as a geometrical property of light -- traces the unresolved structures in the origin of emission. The successful launch and operation of the Imaging X-ray Polarimetry Explorer (IXPE) \citep{Weisskopf2022} has recently brought fresh air into the often overlooked field of X-ray polarimetry and consequently enabled the opening of a new observational window for measuring polarization in the 2--8 keV band for virtually all X-ray sources, including AGNs. 

During the first year and a half of IXPE observations, a few brightest type-1 AGNs (viewed pole on, directly into the nucleus of the axially symmetric accreting structure) were observed: MCG 05-23-16 \citep[$< 3.2\%$ of polarization fraction in 2--8 keV at a 99\% confidence level,][]{Marinucci2022, Tagliacozzo2023}, NGC 4151 \citep[$4.9 \pm 1.1$ \% in 2--8 keV at a 68\% confidence level,][]{Gianolli2023}, and IC 4329A \citep[$3.3 \pm 1.9$ \% in 2--8 keV at a 90\% confidence level,][]{Ingram2023}. In addition, the brightest prototypical type-2 AGN (viewed edge on through the parsec-scale obscuring equatorial material), the Circinus Galaxy, was observed \citep[$20.0 \pm 3.8$ \% in 2--6 keV at a 68\% confidence level,][]{Ursini2023}. Because more fascinating observational results in X-ray polarimetry are yet to come in this decade through the IXPE mission, or its successors [e.g. the XL-Calibur experiment in 15--80 keV \citep{Abarr2021} or the eXTP mission in 2-12 keV \citep{Zhang2016, Zhang2019}], it is timely to put effort simultaneously in the theory and numerical modelling, in order to interpret these results in a meaningful manner.

In this paper, we present an updated and revised summary of the X-ray spectro-polarimetric models of the equatorial parsec-scale AGN components (i.e. the dusty tori) using the {\tt STOKES} code \citep{Goosmann2007,Marin2012,Marin2015,Marin2018}. {\tt STOKES} is a Monte Carlo (MC) radiative transfer code that was originally developed to trace polarization signatures of absorption and scattering in the optical and UV bands, but more recently has been adapted to the X-rays. It includes all important line and continuum processes in the X-ray band and thus forms an ideal tool to study the interaction of light and matter in the central regions of AGNs. {\tt STOKES} is a successor of early MC X-ray spectro-polarimetric simulators of the dusty torus, such as \cite{Ghisellini1994, Ratheesh2021}, and an equal partner to other contemporary MC spectro-polarimetric codes for the toroidal AGN structures, such as the {\tt SKIRT} code \citep{Baes2011, Baes2015, Camps2015, Camps2020} that recently expanded to the X-rays \citep{Meulen2023}. Although in the X-ray band the polarization treatment has not been thoroughly tested yet with {\tt SKIRT}.

We will study solely the equatorial reprocessing of low polarized [up to a few \% \citep{Beheshtipour2017, Tamborra2018, Marinucci2018, Krawczynski2022b, Ursini2022, Krawczynski2022, Marinucci2022, Tagliacozzo2023, Ingram2023, Gianolli2023}] X-ray power-law originating in the central accretion engine that is attributed to hot coronal plasma nearby the accretion disc \citep[see e.g.][]{Haardt1991, Haardt1993}. We aim at those AGNs that do not have strong winds or jets in the polar directions, i.e. the radio-quiet AGNs. Without the addition of polar scatterers, this simple scenario can already serve as a good approximation to a full radio-quiet type-2 AGN model, where equatorial scattering off the dusty torus is of prime importance on parsec scales \cite{Marin2018b}. The argumentation from IR spectroscopy and X-ray changing-look AGNs leads to the clumpy structure of the equatorial scatterer \citep{Nenkova2002, Risaliti2002, Matt2003, Tristram2007}. Tori of non-uniform density were spectro-polarimetrically modeled and compared to smooth tori by {\tt STOKES} as well as other codes mainly in the optical and UV band \citep[see e.g.][]{Nenkova2008a, Nenkova2008b, Marin2015, He2016} and in X-ray thoroughly from a spectroscopic view \citep{Liu2014, Furui2016, Buchner2019, Meulen2023}. Even though pioneering attempts to model the X-ray polarization of clumpy parsec-scale components of AGNs were done by \cite{Marin2012b, Marin2013, Marin2015b}, we restrict ourselves to a homogenous medium in this paper and leave a more detailed study of clumpy equatorial (or polar) scatterers for future works.

A homogenous static torus can already largely affect the polarization outcome depending on its size, location and shape, which are all poorly constrained from both observational and theoretical studies \citep{Fritz2006, Suganuma2006, Namekata2014, Siebenmorgen2015, Buchner2019, Ricci2023}. Although it has not been thouroughly investigated in the X-ray band, MC simulations in the near-IR, optical and UV bands suggest changing polarization due to changing geometry of the equatorial homogenous dusty media on sub-parsec and parsec scales \citep{Kartje1995, Wolf1999, Watanabe2003, Goosmann2007}. We expand these studies into X-rays by comparing wedge-shaped tori with donut-shaped tori of elliptical profiles with various eccentricities for broad range of half-opening angles, observer's inclinations, inner and outer boundaries, ionization levels and densities of the medium, using {\tt STOKES}. We also study the effects of changing the  primary radiation, that is approximated using a point-source power-law. We do not model any timing properties as detailed time-resolved X-ray polarimetry is not of primary interest for the observations of obscuring matter around accretors. We do not assign any velocity to the equatorial material, representing e.g. an outflowing wind from an accretion disc, although it is known to change the X-ray polarimetric properties due to aberation effects \citep[see e.g.][]{Poutanen2023}. Any high-resolution radiative transfer treatment of e.g. the spectral lines is also out of scope of this study, although we note the ambitious attempt of estimating the X-ray polarization fraction from the warm absorber flow in AGNs through detailed radio-hydro-dynamical computations by \cite{Dorodnitsyn2010, Dorodnitsyn2011} with the emphasis on possible polarization gained through resonance lines in a particular large-scale accretion scenario.

Besides the strongly accreting supermassive black holes, a few accreting stellar-mass black holes have been already observed by the IXPE mission \citep{Krawczynski2022, Veledina2023, Podgorny2023b, Rawat2023, Kushwaha2023, Ratheesh2023, Cavero2023}. These are located inside X-ray binary systems (XRB) and, apart from their shorter physical sizes, show remarkably similar properties to AGNs, including the accretion disc structure, the hot X-ray emitting coronae above, below or within the disc, and the formation of jets surrounded by ionization cones in the polar directions. Even though the dusty torus in its grandeur is forming only around AGNs when switching to the accretion-efficient state \citep{Schawinski2015}, sometimes a covering equatorial medium in a funnel shape is considered for XRBs or ultra luminous X-ray sources (ULX) \citep[see e.g.][]{Wielgus2016, Narayan2017, Ratheesh2021}. Of direct relevance to our study is the recent discussion in \cite{Veledina2023} on the IXPE observation of Cygnus X-3 XRB ($20.6 \pm 0.3$ \% and $10.4 \pm 0.3$ \% in 2--8 keV at a 68\% confidence level in two subsequent observations separated by $\sim$2 months), where the models of AGN tori described in this paper have been used to interpret the X-ray polarimetric data, alongside new analytical models for single scatterings on the inner surface of a conical funnel. Therefore, we point to the applicability of AGN modelling results to the neighboring field of XRBs and vice versa. Moreover, the Be-transient X-ray pulsar LS V +44 17/RX J0440.9+4431 was recently observed by IXPE \citep[for two subsequent observations $4.4 \pm 0.2$ \% and $4.9 \pm 0.2$ \% in 2--8 keV at a 68\% confidence level,][]{Doroshenko2023} with part of the observed polarization suggested to arise due to a partially ionized equatorial obscurer. Thus, we note potential usefuleness of some of the ideas presented here for the accreting neutron stars and white dwarfs studies, because of the covering outflows present also near such compact objects.

The paper is organized as follows: in Section \ref{methods} we introduce the physical processes, assumptions and numerical methods used. Section \ref{equatorial} provides the broad-range X-ray spectro-polarimetric modelling results of reprocessing of a central power-law emission by axially symmetric homogeneous equatorial obscurer of various types. An example of an application of the MC models to interpretation of the X-ray polarimetric observation of XRB Cygnus X-3 is presented in Section \ref{cygx3_results}. We conclude in Section \ref{conclusions}.

\section{Modelling}\label{methods}

The Monte Carlo computations were made with the {\tt STOKES} code \footnote{\url{www.stokes-program.info}} \citep{Goosmann2007,Marin2012,Marin2015,Marin2018}, version \textit{v2.07} that is suitable for X-rays and that was also used in \cite{Marin2018b, Marin2018c}. The code is ideal for studying polarization properties of light in media where scattering and absorption is the dominant source of opacity. It allows combinations of user-defined 3D regions with a given density, scattering and absorbing structure. All known line and continuum processes in the X-rays, such as photo-electric absorption and multiple Compton down-scattering, are included with the exception of Compton up-scattering, and synchrotron and thermal emission that would need to be corrected for ad hoc (not self-consistently, because the temperature and magnetic fields are not computed) and that should not play a major role in the cold parsec-scale components of AGNs or distant outflows of accreting stellar-mass compact objects.

The emitting region can be located at arbitrary position and any incident polarization state. The user defines its energy distribution and initial polarization state. The 3D geometry is given by the origin of Cartesian coordinates \{$x$, $y$, $z$\}, not taking into account any distortion of space-time by strong gravity. We place the X-ray emitting region representing the central disc-corona system in the center of our coordinates, neglecting its physical size, but assuming that the disc lies in the equatorial $xy$-plane. We assume an isotropic power-law emission with a power-law index $\Gamma$ from the central point that is either unpolarized, $2\%$ polarized (in Section \ref{equatorial}), or $4\%$ polarized (in Section \ref{cygx3_results}). If polarized, we consider the input parallely or perpendicularly polarized with respect to the axis of symmetry of the system, although general-relativistic effects in the vicinity of the central compact object and non-trivial superposition of radiation can result in different polarization angles in-between the two directions we defined \citep[see e.g.][]{Krawczynski2022b, Ursini2022}. Photons are detected by a spherical web of virtual detectors with particular energy binning and azimuthal and polar binning from the central source, allowing symmetrization in mid-plane and around the $z$-axis of the scattering region. An auxiliary routine {\tt ANALYZE} created by the authors of {\tt STOKES} is used in the end to convert the output to conveniently readable files storing the Stokes parameters $I$, $Q$ and $U$, per energy bin and angular bin of each detector. We consider the energy range between 1 and 100 keV with a medium energy resolution of 200 logarithmically spaced bins to be able to discuss all major spectro-polarimetric properties of the re-processed radiation. We will use 20 bins (in Section \ref{equatorial}) or 40 bins (in Section \ref{cygx3_results}) for the observer's inclination $0\degr < i < 90\degr$ measured from the pole and one azimuthal bin, expecting axial symmetry.

\begin{figure}
	\includegraphics[width=1.\columnwidth]{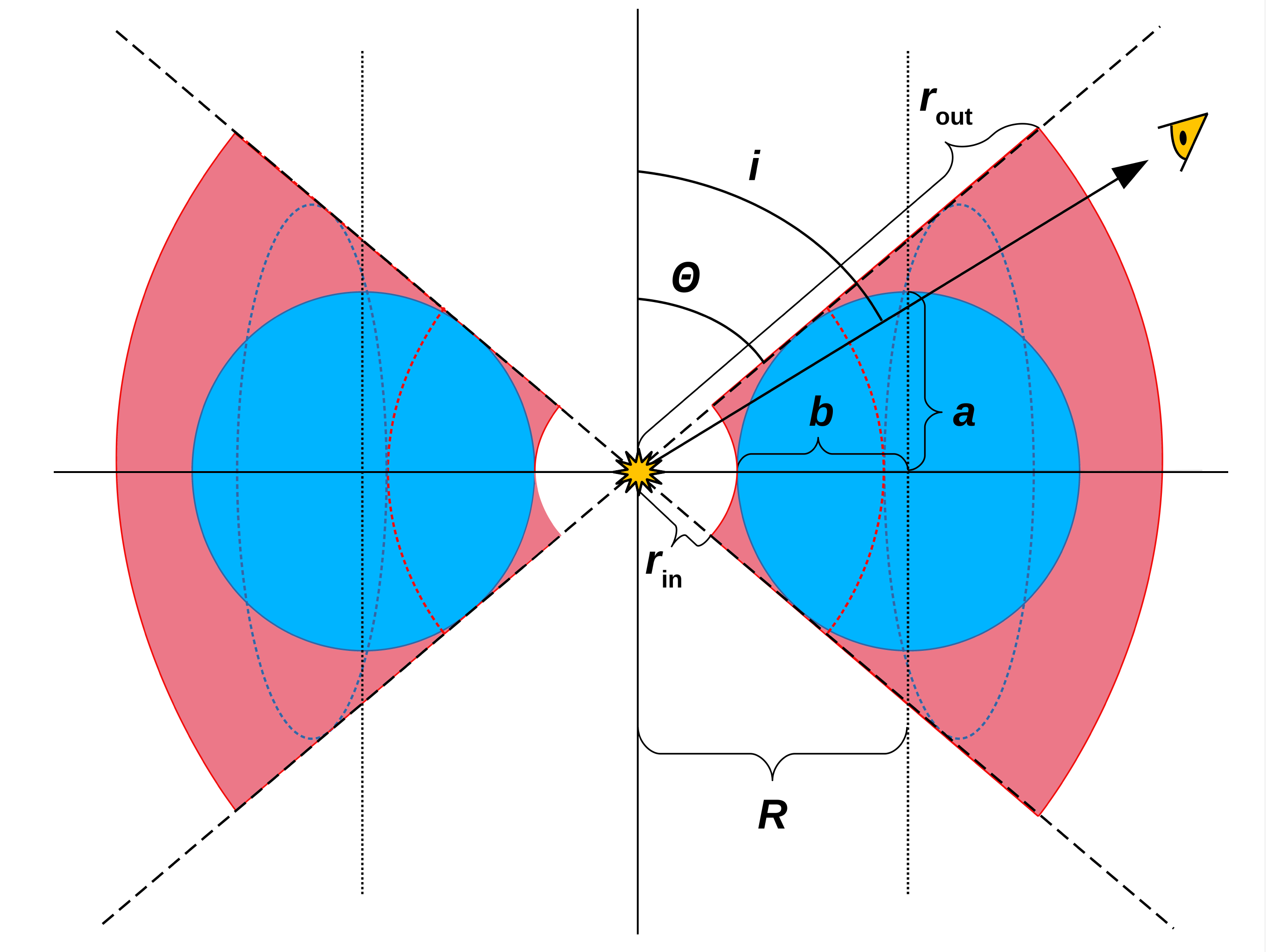}
	\caption{A schematic in the meridional plane of the axially symmetric equatorial reprocessing regions studied. The illuminating disc-corona region is located in the center of the coordinate system, approximated as a point source. In Section \ref{equatorial}, we study reprocessing in various homogenous equatorial media that differ primarily in geometry. Case A (in red) represents the wedge-shaped geometry, Case B (in dotted blue) represents a general elliptical torus (i.e. $a \neq b$), and Case C (in blue) is a special subcase of the latter that represents a circular torus (i.e. $a = b$). A ``type-1'' observer is referred to when $i < \Theta$. A ``type-2'' observer is referred to when $i > \Theta$. See text for details for the rest of the parameters.}
	\label{stokes_mo}
\end{figure}

A schematic representation of the equatorial scatterers studied in the paper is shown in Figure \ref{stokes_mo}, viewed edge-on through the meridional plane. All ``tori'' studied in this paper have uniform densities, are symmetric with respect to the black-hole rotation axis and reside symmetricaly in the equatorial plane around the accretion disc (shrinked to negligible size in {\tt STOKES}). We study
\begin{itemize}
    \item Case A: a wedge-shaped torus defined by $r_\textrm{in}$, $r_\textrm{out}$ and its half-opening angle $\Theta$.
    \item Case B: a true geometrical torus with an elliptical profile given by the parameters $R$, $a$ and $b$.
    \item Case C: a subcase of B representing a torus with circular profile, i.e. when $a = b$.
\end{itemize}
The choices of parametric values for the displayed results are physically and observationally motivated and we refer the reader to \cite{Goosmann2007, Marin2018b, Marin2018c} for further details. We always directly compare the wedge-shaped tori with $r_\textrm{in}$ fixed to the inner-most circle of the elliptical tori to study the effects of concave or convex inner surfaces. For Case A, $r_\textrm{out}$ is left as a free parameter.

Let us first describe the geometrical comparisons in Section \ref{equatorial}, where we only trace the reprocessed radiation, i.e. photons that scattered at least once before reaching the observer (this is not the case for Section \ref{cygx3_results} where we include all photons). For Case B, we set $b$ relative to $R$ and study various eccentricities from prolate to oblate tori: $b = R/8$, $b = R/4$, $b = R/2$, $b = 3R/4$, where $R = b + r_\textrm{in}$ for a given $r_\textrm{in}$. Thus, through changing the eccentricity, we also trace the effects of changing curvature of the inner side. For a given half-opening angle $\Theta$ measured from the pole, we then have in Case B the parameter $a$ self-consistently computed via :
\begin{equation}\label{size_trafo}
    a = \tan(\frac{\pi}{2} - \Theta)\sqrt{R^2-b^2} \textrm{ .}
\end{equation}
Case C is denoted as a special case of an elliptical torus, because we will use it as a referential torus shape for future model comparison (Podgorn\'y et al. in prep.). Also Case C has $a = b$ computed in a different way than the Case B tori, allowing direct relative size comparisons in the resulting polarization within one geometrical architecture.
For case C :
\begin{equation}\label{opening_angle}
    a = b = r_\textrm{in}\frac{\cos{\Theta}}{1-\cos{\Theta}} \textrm{ ,}
\end{equation}
where again $r_\textrm{in}$ is fixed. In this way, for a given $\Theta$, when comparing Case C to Cases B with any $b/R$ ratio, the torus radii $R$ are different, but the inner boundary remains the same between the two cases. Case B would equal Case C, if $b/R=\cos{\Theta}$. We always directly compare the geometrical shapes with the same $\Theta$ and observer's inclination $i$ (representing a type-1 view, if $i < \Theta$, and type-2 view, if $i > \Theta$).

In Section \ref{cygx3_results} we also allow to register photons that are reaching the observer directly from the central emission, without any scattering, in order to study the transparency effects and provide a rather realistic example in application to XRBs, compared to the theoretical reprocessing-only discussion in Section \ref{equatorial}. In addition, for geometrical comparisons in Section \ref{cygx3_results}, we use an alternative style for plotting the results, in order to provide a different perspective and direct comparisons with the analytical model presented in \cite{Veledina2023} and MC simulations from \cite{Ghisellini1994} adapted to study the torus in Circinus Galaxy in \cite{Ursini2023}. Otherwise the geometrical setup is the same, except we loosen the condition of a fixed $r_\textrm{in}$ and we fix $R$ instead to study the effects due to various inner edges. In this way, for Case B, we have $b$ again given by the ratios $b = R/8$, $b = R/4$, $b = 3R/4$ and $a$ consistently computed from Equation \ref{size_trafo}. The wedge-shaped Case A has $r_\textrm{in}$ computed from $r_\textrm{in} = R - b$, where $b = a$ is from a circular torus with $b = R\cos{\Theta}$.

For all displayed results, the composition of the axially symmetric equatorial scattering regions assumes the same atomic species used for \cite{Marin2018b, Marin2018c} and the solar abundance from \cite{Asplund2005} with $A_\mathrm{Fe} = 1.0$. The uniform neutral hydrogen density $n_\textrm{H}$ is given by the total column density $N_\textrm{H} = n_\textrm{H}L$, where $L$ is the size of the scattering region between the center and an equatorial observer. Apart from the Case A, $N_\textrm{H}$ of the scattering region in the line of sight varies with inclination, but the photons that reach the observer can reprocess everywhere in the model and change direction according to the scattering laws. Moreover, they are allowed to escape the material and be reprocessed again in other part of the torus, allowing self-irradiation effects. We also define the optical depth for scattering on free electrons $\tau_\textrm{e} = \sigma_\textrm{T} n_\textrm{e} L$ via the standard Thomson opacity $\sigma_\textrm{T}$ and free electron density $n_\textrm{e}$, which allows to trace effects of partial ionization. The only exception in composition is the application example discussed in Section \ref{cygx3_results}. There, for the case of Cygnus X-3 we, expect the circumnuclear outflows to be orginating from stellar winds of the companion, which is a Wolf-Rayet star with severe hydrogen depletion in the outer atmospheric layers due to its past stellar evolution \citep{Fender1999, Kallman2019}. Thus, we will keep the abundances and metalicities, and only remove the most abundant neutral hydrogen, providing rather the $N_\textrm{He}$ values, which are more informative as neutral helium is then the most abundant element. In this way, we can asses the effects of changing chemical composition in the system.

We define the linear polarization degree $p$ and polarization position angle 
$\Psi$ using the Stokes parameters $I$, $Q$ and $U$
\begin{equation}\label{ppsidef}
	\begin{aligned}
		p &= \dfrac{\sqrt{Q^2+U^2}}{I} \\
		\Psi &= \dfrac{1}{2}\textrm{\space}\arctan_2\left(\dfrac{U}{Q}\right)   \textrm{ ,}
	\end{aligned}
\end{equation}
where $\arctan_2$ denotes the quadrant-preserving inverse of a tangent function and $\Psi = 0$ means that the polarization vector is oriented \textit{parallel} to the system axis of symmetry projected to the polarization plane. $\Psi$ increases in the counter-clockwise direction from the point of view of an incoming photon. Due to symmetry, the polarization genesis is such that the observed photons are polarized on average parallel or perpendicular to the axis of symmetry, thus we will use the notation of positive or negative polarization degree $p$ for such resulting $\Psi$, respectively.

\section{Results}\label{equatorial}

Throughout this section (in contrast to Section \ref{cygx3_results}), we study reprocessed light only, i.e. we do not register photons that did not scatter at least once before reaching the observer. We set $r_\textrm{in} = 0.05$, representing a fraction of a parsec for the outer accretion disc edge or the inner torus boundary in AGNs. Nonetheless, the units are arbitrary and the results can be used for orders of magnitude lighter objects, because for polarization genesis in this approach only the relative sizes inside the system matter. A few other values of $r_\textrm{in}$ were tried with generally lower impact on the polarization results than other parameters (see also Section \ref{cygx3_results} for a fixed $R$ and changing geometry with respect to $r_\textrm{in}$), but the effects of changing distance to the center $r_\textrm{in}$ relative to other proportions should be more thoroughly investigated in the future. The outer boundary of Case A is set to $r_\textrm{out} = 10$, representing the outermost torus boundary in parsecs. Unlike the inner edge, the outer edge distance and geometrical shape plays little role on the emergent polarization, as long as the column densities of the entire scattering and absorption region are kept constant and high enough. Because in that case the first orders of scattering happening on the inner walls are determinative for the total geometry of scattering of unabsorbed photons. In order to first discuss the geometry and composition effects of the scattering regions, we display the results for a generic example of the primary radiation with $\Gamma = 2$ and 2\% parallel polarization with respect to the axis of symmetry, if not stated otherwise. We tested three cases of ionization, referring to them as low, moderate and high ionization level ($\tau_\textrm{e} = 0.07, 0.7, 7$, respectively, for $N_\textrm{H} = 10^{25}  \textrm{ cm}^{-2}$ and $\tau_\textrm{e} = 0.007, 0.07, 0.7$, respectively, for $N_\textrm{H} = 10^{24}  \textrm{ cm}^{-2}$). In this way we study a range of ionizations from nearly fully ionized medium to nearly neutral medium, corrected for density change of neutral atomic species.

\begin{figure*}
	\includegraphics[width=2.07\columnwidth]{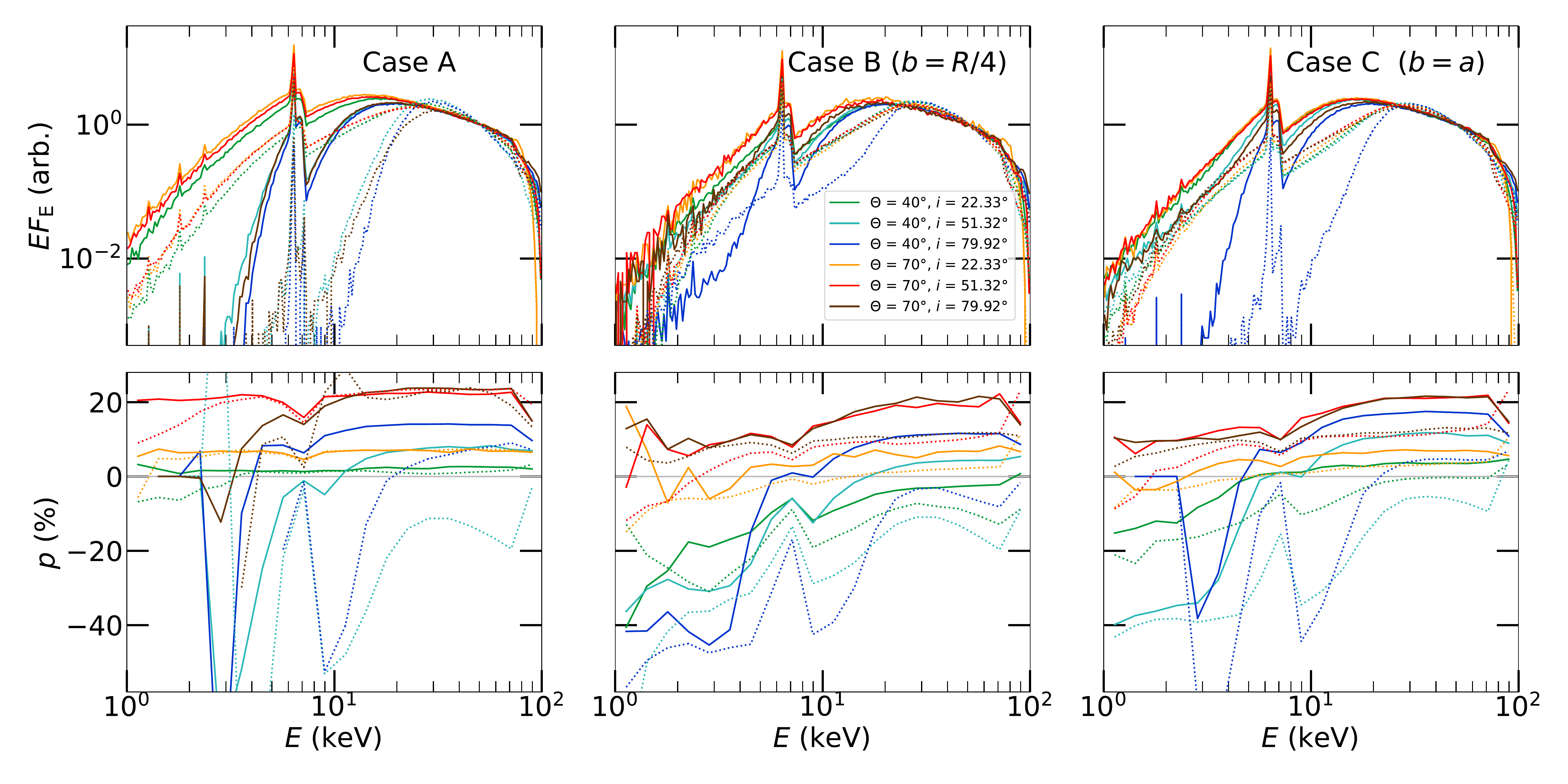}
	\caption{Top: the spectra, $EF_\mathrm{E}$, normalized to value at 50 keV. Bottom: the corresponding polarization degree, $p$, versus energy. We show the Cases A, B (for $b = R/4$) and C from left to right. Various inclinations $i$ and half-opening angles $\Theta$ for both type-1 and type-2 observers are shown in the color code. We show the torus column densities $N_\textrm{H} = 10^{24}  \textrm{ cm}^{-2}$ and $\tau_\textrm{e} = 0.7$ (solid lines) and $N_\textrm{H} = 10^{25}  \textrm{ cm}^{-2}$ and $\tau_\textrm{e} = 7$ (dotted lines), i.e. the highly ionized cases. The remaining parameters are $r_\textrm{in} = 0.05$ (Case A), $R_\textrm{out} = 10$ (Case A) and the primary input was set for $\Gamma = 2$ and $p_0 = 2$\% for all displayed cases. The detailed parametric dependencies are shown in following figures when integrated in energy. The fluorescent spectral lines are depolarizing on top of the continuum. High photo-absorption (stronger for higher densities and at softer energies) causes higher perpendicular polarization contribution to the total emission. The details are geometry dependent. In the bottom-left and bottom-right panels, the high polarization at 3--5 keV in blue and turquoise cases reaches 75\% to 100\%. We reduced the $y$-axis range, because any such values are loosely determined due to simulation precision that suffers from high absorption, hence low photon statistics in these bins.}
	\label{energy_dependent}
\end{figure*}

Figure \ref{energy_dependent} shows the energy dependence of flux and polarization degree for a few examples from all cases studied, which are further directly compared to each other integrated in energy in order to study other parametric dependencies in detail. Compared to the total flux, the polarization degree is binned to 10 neighboring bins in energy for better MC statistics in the figure, but the $Q$ and $U$ parameters were computed in the same resolution as $I$. We clearly see depolarization in the emission lines, which is expected \citep[see e.g. ][]{Goosmann2011, Marin2018b, Marin2018c}, compared to polarization gained through Compton scattering. The drop of polarization in fluorescent lines is in detail dependent on each line's formation mechanism, and also on the scattering and absorption conditions that can add polarization to unpolarized photons originating in emission lines elsewhere, such as the accretion disc, winds, or other part of the dusty torus, from which the emission line photon may escape to be reprocessed again. Therefore, we will not aim for a detail study of these effects due to low photon statistics and complexity of the problem. Especially at energies lower than 3.5 keV where many emission lines are present and where for lower densities the flux is low due to energy-dependent opacities (see top panels of Figure \ref{energy_dependent}). We also note there could be additional increase of polarization via resonant lines, which are treated by the {\tt STOKES} code, but in terms of spatial and energy resolution not in such detail as in the radiative transfer estimates for warm absorbers by \cite{Dorodnitsyn2010, Dorodnitsyn2011}. These studies claim the polarization produced by means of resonant lines below 2 keV to be as important as Compton scatterings. We will later compare the energy ranges 3.5--6 keV and 30--60 keV to discuss pure continuum results in middle and hard X-rays, while the first band is especially important for IXPE operating in 2--8 keV.

The flux, and consequently the polarized flux, which is an important quantity for studying detectability of polarization, is energy-dependent also due to obscuration effects, i.e. we see higher absorption for lower half-opening angles and higher inclination angles. Although absorption in general enhances polarization in slab-like problems \citep[e.g. for the atmospheres of accretion discs in plane-parallel approximation][]{LightmanShapiro1975, LoskutovSobolev1979, LoskutovSobolev1981}, in toroidal geometries this is not always expected. The detailed reasoning is quite complex, because relative polarized flux contributions from different regions of the scatterer need to be considered, including the dominant polarization angle from each part. We see highly polarized reflection from the inner side of the funnel (depending on the ionization level), the flux contribution from each section is inclination and geometry dependent, and reprocessing occurs at various volumes, depending on the optical depth (in turn dependent on the density) of the medium. Mirror symmetry in meridional plane leads to either parallel or perpendicular orientation of polarization from a solid angle subtended by each meridionally symmetric region of the torus with a given relative flux and polarization fraction contribution, which has a consequence on the net polarization. In scattering-dominated media the following geometrical explanation holds for the emergent polarization angle: if the dominant scattering plane is equatorial, i.e. the observed scattering medium is rather equatorially elongated, the expected polarization is parallel to the axis of symmetry; if it is the meridional plane, then the net polarization arising due to the dominant single-scattering angle will be perpendicular to the principal axis \citep[see e.g.][]{Kartje1995, Goosmann2007, Schnittman2010, Krawczynski2022}. Then, higher overall flux absorption may also cause depolarization due to mutually perpendicular polarization vectors of light rays originating in different regions and reaching the observer. In the bottom panels of Figure \ref{energy_dependent}, we see a general trend of increasing perpendicular polarization towards lower energies at soft X-rays due to relatively high flux contribution from scattering at the further side of the torus, which then reaches the observer with a dominant single-scattering angle in the meridional plane. Hence with high perpendicular polarization gained through Compton scattering ``periscope'' effect, because the resulting polarization is always perpendicular to the dominant plane of scattering. The opposite is true for hard X-rays, especially for low column densities when partial transparency allows larger visibility angles of the inner illuminated walls (not just the back side), and a dominant parallel angle due to scatterings in parts of the medium that are on the left and right sides from observer's point of view (together forming a larger polarized flux contribution through the total observable solid angle) and rather elongated in the equatorial plane. Their lower obscuration (i.e. the opacity drop at hard X-rays due to photo-electric absorption optical depths being roughly $\sim E^{-3}$ dependent compared to nearly constant Compton scattering optical depths with energy) causes higher polarization with parallel orientation in most of the lower density cases and high half-opening angles -- creating rather equatorial ring scattering environments than vertically extended walls. Thus at hard X-rays, in the cases were the net emergent polarization angle is parallel, we rather expect depolarization for higher absorption (by higher uniform density), because the higher obscuration emphasizes observed regions with the competing perpendicular polarization angle. The results are geometry dependent and although the two elliptical tori shown, i.e. Case B (for $b = R/4$) and Case C ($b = a$), are rather similar, the wedge-shaped Case A preferentially shows parallel orientations, even at lower energies. This is logical, because the column density in the line of sight towards the primary source is inclination independent (a step function in between type-1 and type-2 viewing angles) for wedge-shaped tori, unlike the elliptical tori, where the transition in line-of-sight column density is more gradual. Therefore, we expect different contribution of light scattered in different volumes along the entire radial extent near the grazing angles. They also undergo different geometry of scattering on the inner side, due to a change in curvature sign of the inner surface with respect to the center of emission between Case A and Cases B and C.

\begin{figure*}
	\includegraphics[width=2.\columnwidth]{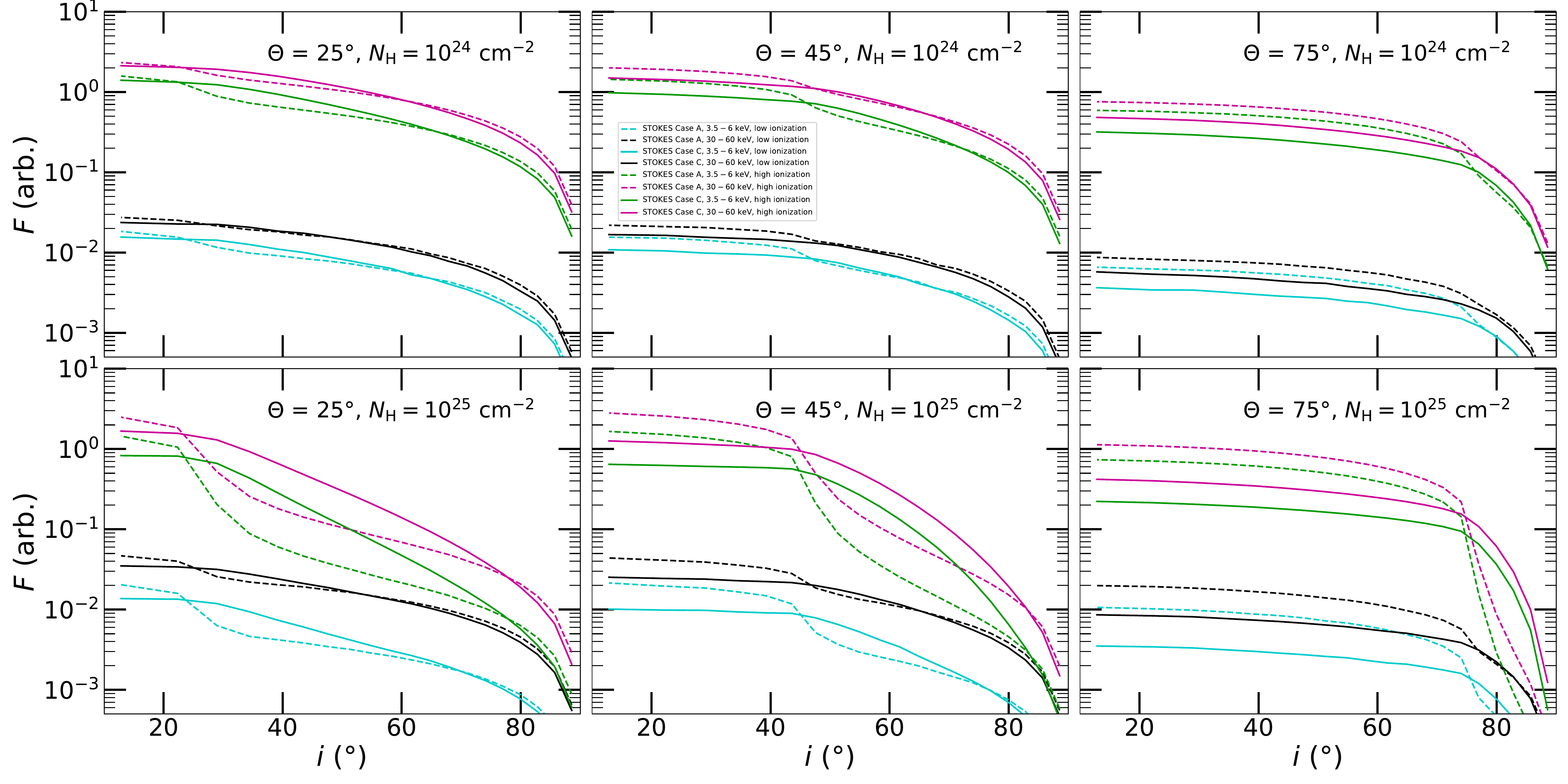}
	\caption{Case A (dashed lines) compared to Case C (solid lines): we show the energy-integrated flux, $F$, versus inclination, $i$, for $\Theta = 25\degr$ (left), $\Theta = 45\degr$ (middle), and $\Theta = 75\degr$ (right) and for $N_\textrm{H} = 10^{24}  \textrm{ cm}^{-2}$ (top) and $N_\textrm{H} = 10^{25}  \textrm{ cm}^{-2}$ (bottom). We show the results for \textit{low} level of ionization, i.e. $\tau_\textrm{e} = 0.007$ for $N_\textrm{H} = 10^{24}  \textrm{ cm}^{-2}$ and $\tau_\textrm{e} = 0.07$ for $N_\textrm{H} = 10^{25}  \textrm{ cm}^{-2}$, integrated in 3.5--6 keV (blue) and 30--60 keV (black), as well as for \textit{high} level of ionization, i.e. $\tau_\textrm{e} = 0.7$ for $N_\textrm{H} = 10^{24}  \textrm{ cm}^{-2}$ and $\tau_\textrm{e} = 7$ for $N_\textrm{H} = 10^{25}  \textrm{ cm}^{-2}$, integrated in 3.5--6 keV (green) and 30--60 keV (magenta). The remaining parameters are $r_\textrm{in} = 0.05$ (Case A), $R_\textrm{out} = 10$ (Case A) and the primary input was set to $\Gamma = 2$ and $p_0 = 2$\% for all displayed cases. The ionization level impacts reflectivity of the medium, while the geometry and density impacts the shape of the flux drop with inclination. For wedge-shaped tori (Case A), the total flux decreases relatively steeply below the grazing angle due to sudden increase of line-of-sight column density with inclination compared to the circular torus (Case C).}
	\label{AvsC_Imue}
\end{figure*}

We will now examine the results in more detail by integrating the spectra in energy in ranges where spectral lines do not contribute. We will include the discussion of partial ionization effects that were not shown in Figure \ref{energy_dependent}. We will first compare the wedge-shaped tori with circular tori. Figure \ref{AvsC_Imue} shows the relative flux drop with inclination for Case A vs. Case C for various densities, ionizations and half-opening angles. The shape of the circular torus, with a concave upper and lower boundary in the meridional plane compared to the straight wedge-shaped torus, clearly impacts the received flux through gradually decreasing transparency when changing the view from the grazing angle towards the equator. Whereas the wedge-shape tori provide much steeper flux drops with a sharp column density switch in the line of sight just below the grazing angle. This effect is more important for higher densities of the matter and is rather ionization independent, although higher ionization means higher flux due to enhanced reflection. Since we show the results in continuum dominated energies, the relative flux change is rather energy-independent.

\begin{figure*}
	\includegraphics[width=2.\columnwidth]{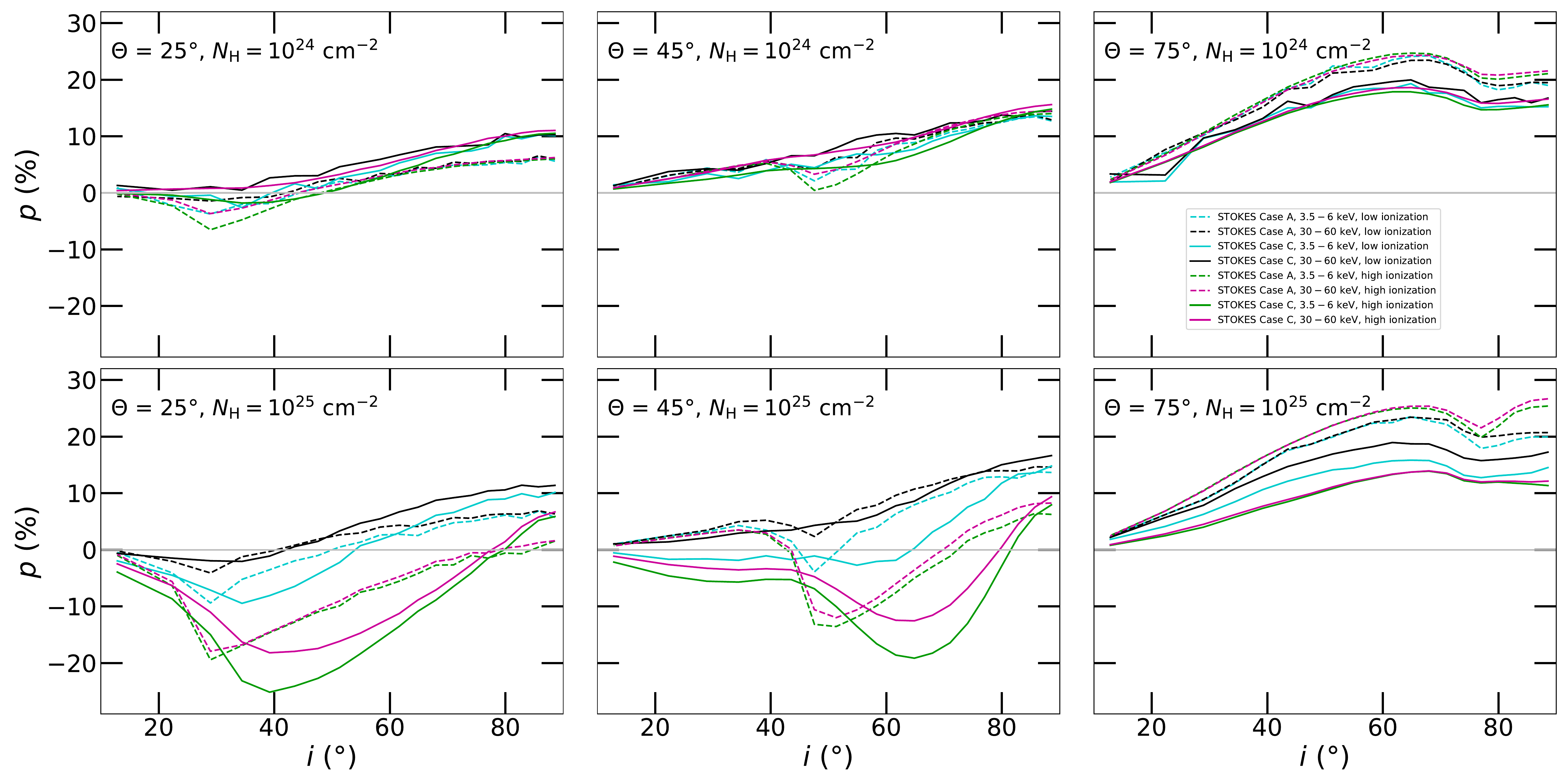}
	\caption{The same as in Figure \ref{AvsC_Imue}, but energy-averaged polarization degree, $p$, versus inclination $i$ is shown. The reprocessed polarization can reach tens of \% and geometrical details are important. For lower densities, the prevalent polarization orientation is parallel with the axis of symmetry. Circular tori (Case C) then in some cases of high obscuration lead to a resulting perpendicular polarization.}
	\label{AvsC_pmue}
\end{figure*}

\begin{figure*}
	\includegraphics[width=2.\columnwidth]{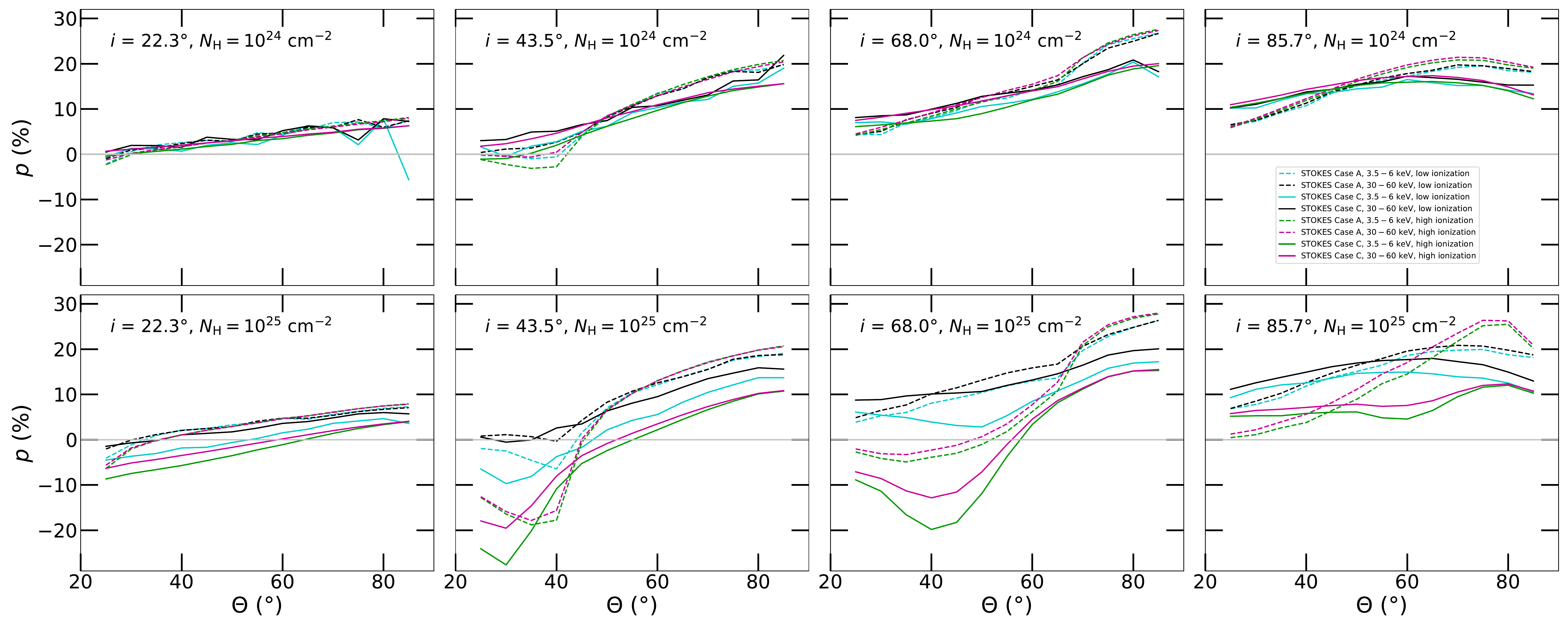}
	\caption{The same as in Figure \ref{AvsC_pmue}, but energy-averaged polarization degree, $p$, versus half-opening angle $\Theta$ is shown for $i = 22.3\degr$, $i = 43.5\degr$, $i = 68.0\degr$, and $i = 85.7\degr$ (left to right). In different view of the results, we see the polarization dependency on half-opening angle: the smaller the vertical extension of the torus is, the more equatorial reprocessing we witness and the higher and clearer the parallel polarization contribution is, especially for low column densities.}
	\label{AvsC_ptheta}
\end{figure*}

Figures \ref{AvsC_pmue} and \ref{AvsC_ptheta} show the polarization comparison for the same configurations with respect to various inclination and half-opening angle combinations. We see gradually less distinct regimes while changing $i$ from minimum to maximum, and more distinct regimes in $\Theta$ -- that is for the choice of plotting and number of panels. The column density clearly affects the dominant symmetry that results in net observed parallel or perpendicular polarization angle. For lower densities we obtain almost exclusively parallel polarization reaching from unpolarized emission to $\sim 20$\% polarized emission for high inclinations and high half-opening angles. This is due to different angular flux distribution, i.e. the partial transparency is allowing the observer to see more reprocessed light from the left and right sides of the tori than from the far side in observer's point of view, compared to the case of higher densities. Since the front side is illuminated from an opposite side, the flux contribution is low, as almost no radiation passes through the torus in the type-2 viewing angles. Even though we forbid here the registration of zero-scattered photons, the results for obscured viewing angles are not much different when including them, which we did in Section \ref{cygx3_results}. For higher densities, the same scenario is true for high inclinations and high half-opening angles. But when we obscure the central source with low half-opening angles, the reflection on the further side off the inner walls of such funnels prevails and we see a dominant perpendicular orientation with average polarization reaching up to $\sim 25$\%. For this regime, we see depolarization for lower densities, as the escaped photons scatter in larger volumes of the equatorial medium, thus on average come through higher diversity of directions and the overall symmetry is less broken. The most favourable obscuration scenario for perpendicular polarization genesis through on average perpendicular reflection in the meridional plane, i.e. the highest relative contribution regarding flux and polarization, happens at moderate inclinations, from $\sim 40\degr$ to $\sim 70\degr$ between $25\degr$ and $50\degr$ half opening angles. We refer the reader to Section \ref{cygx3_results} where the dependency on $i$ and $\Theta$ is viewed from a different, perhaps more complete perspective for one application example. 

The continuum energy-dependence traces the amount of photo-electric absorption with respect to scatterings. Apart from some irregular contribution of ionization edges (less present in the two energy bands shown), we have higher photo-electric absorption in the 3.5--6 keV band compared to 30--60 keV. This causes an opposite dependence of polarization degree \textit{on energy} with respect to the two dominant regimes appearing in the resulting polarization angle. However, the level of ionization is more important than the energy band chosen towards the resulting polarization fraction, which is seen on both figures. More prominently for higher densities, as the density affects the energy-dependent ratios between bound-free and free-free absorption opacities and scattering opacities on free electrons. The geometrical shape is intermediately important compared to changes in $p$ with energy and ionization. Even though we have the inner radius $r_\textrm{in}$ fixed here when comparing the Case A with Case C, the curvature of the inner boundary and shape of the top or bottom surface with respect to the equatorial plane is determinative for the net geometry of scattering. We thus obtain similar conclusions to the optical and UV study in \cite{Goosmann2007}, which found non-negligible sensitivity of polarization to the details of the geometry, similar half-opening angle and inclination effects on the polarization angle, and polarization degree forming a peak with inclination, i.e. showing a non-monotonous behavior. The results are also similar, although not equivalent, to previous MC studies of equatorial X-ray polarization reprocessing in \cite{Ghisellini1994, Goosmann2011, Ratheesh2021, Ursini2023}. The works of \cite{Ratheesh2021, Ursini2023}, departing from \cite{Ghisellini1994}, stated however only perpendicular orientation of the resulting polarization vector with respect to the axis of symmetry. And even though \cite{Goosmann2011} used the same code, the exact computational conditions, e.g. regarding the geometry and column densities, resulting in different energy dependence, were not the same and they were simulated in particular for the nucleus of NGC 1068, while here we provide much broader parametric overview. One should also compare the results only for the type-2 viewing angles, as in this section we do not register direct photons, only the reprocessed ones.

\begin{figure*}
	\includegraphics[width=2.\columnwidth]{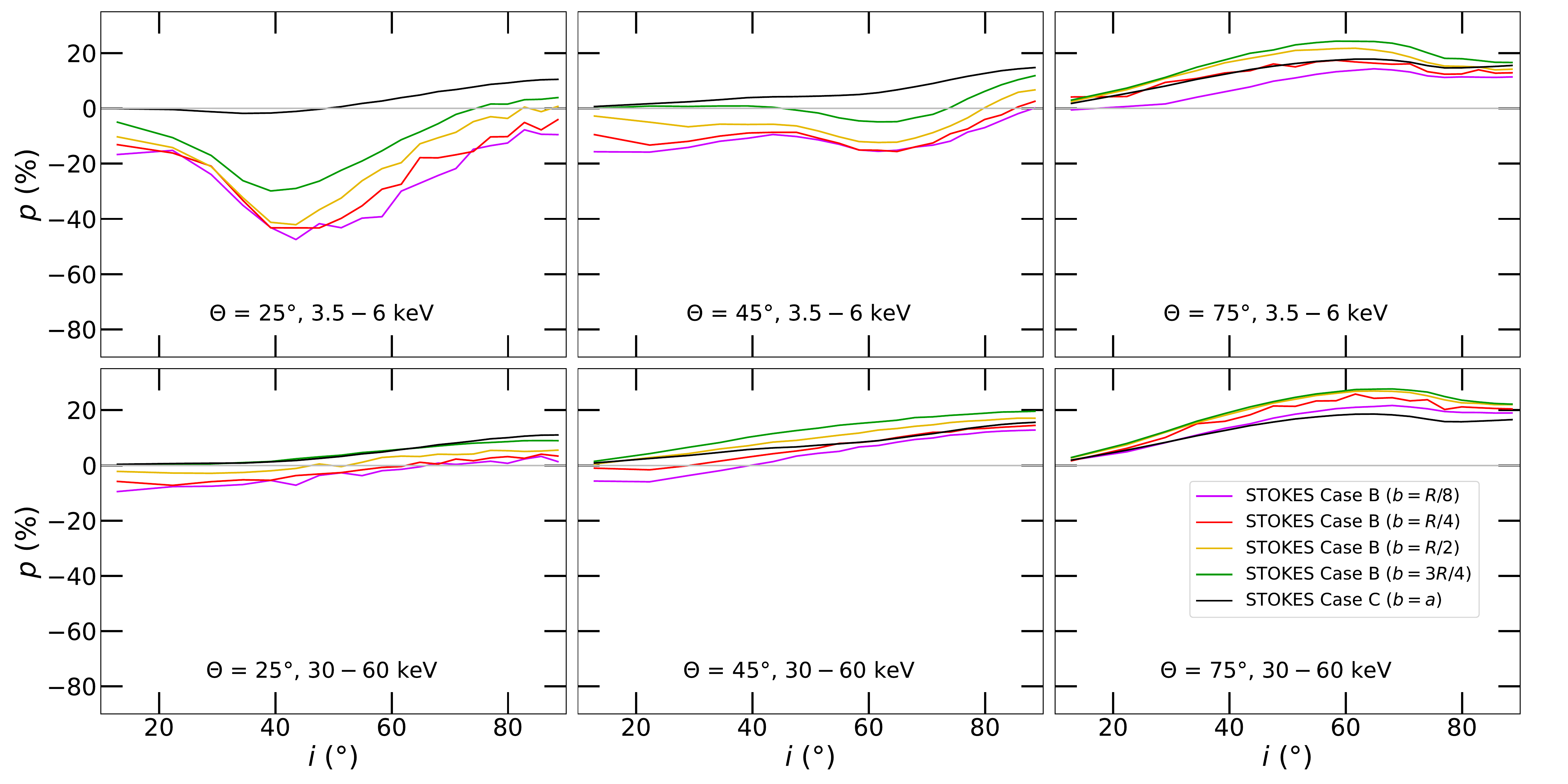}
	\caption{Case B for various ellipse eccentricities compared to the circular Case C in energy-averaged polarization degree, $p$, versus inclination $i$ for $\Theta = 25\degr$ (left), $\Theta = 45\degr$ (middle), and $\Theta = 75\degr$ (right). We show the results integrated in 3.5--6 keV (top) and 30--60 keV (bottom) and for $N_\mathrm{H} = 10^{24} \textrm{ cm}^{-2}$. The black curve corresponds to the Case C ($b = a$). The magenta, red, yellow, green curves correspond to the Case B for $b = R/8$, $b = R/4$, $b = R/2$, $b = 3R/4$, respectively. The results are shown for \textit{high} level of ionization, i.e. $\tau_\textrm{e} = 0.7$, see corresponding Figure \ref{BvsC_pmue24_low_io} for low ionization. The primary input was set to $\Gamma = 2$ and $p_0 = 2$\% for all displayed cases. Due to the adopted definitions, Case C does not always represent the middle case between the more general elliptical Cases B. The more oblate the elliptical scattering structure is, the stronger parallel polarization contribution is and the closer to the wedge-shaped result the net polarization gets.}
	\label{BvsC_pmue24}
\end{figure*}

\begin{figure*}
	\includegraphics[width=2.\columnwidth]{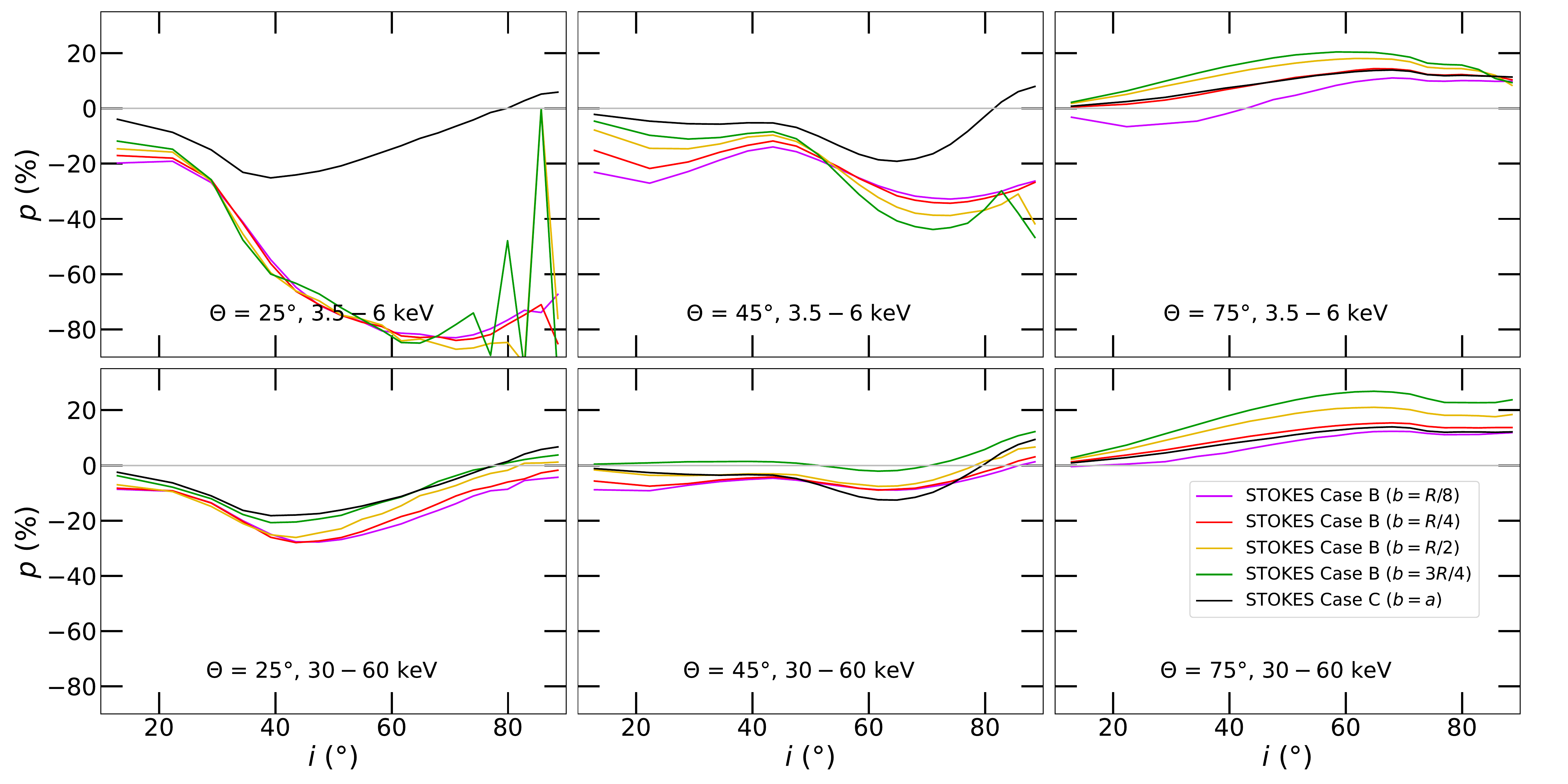}
	\caption{The same as in Figure \ref{BvsC_pmue24}, but for $N_\mathrm{H} = 10^{25} \textrm{ cm}^{-2}$. The results are shown for \textit{high} level of ionization, i.e. $\tau_\textrm{e} = 7$, see corresponding Figure \ref{BvsC_pmue25_low_io} for low ionization. For high enough densities and soft energies, the dependency on eccentricity is reversed for low half-opening angles and high inclinations.}
	\label{BvsC_pmue25}
\end{figure*}

\begin{figure*}
	\includegraphics[width=2.\columnwidth]{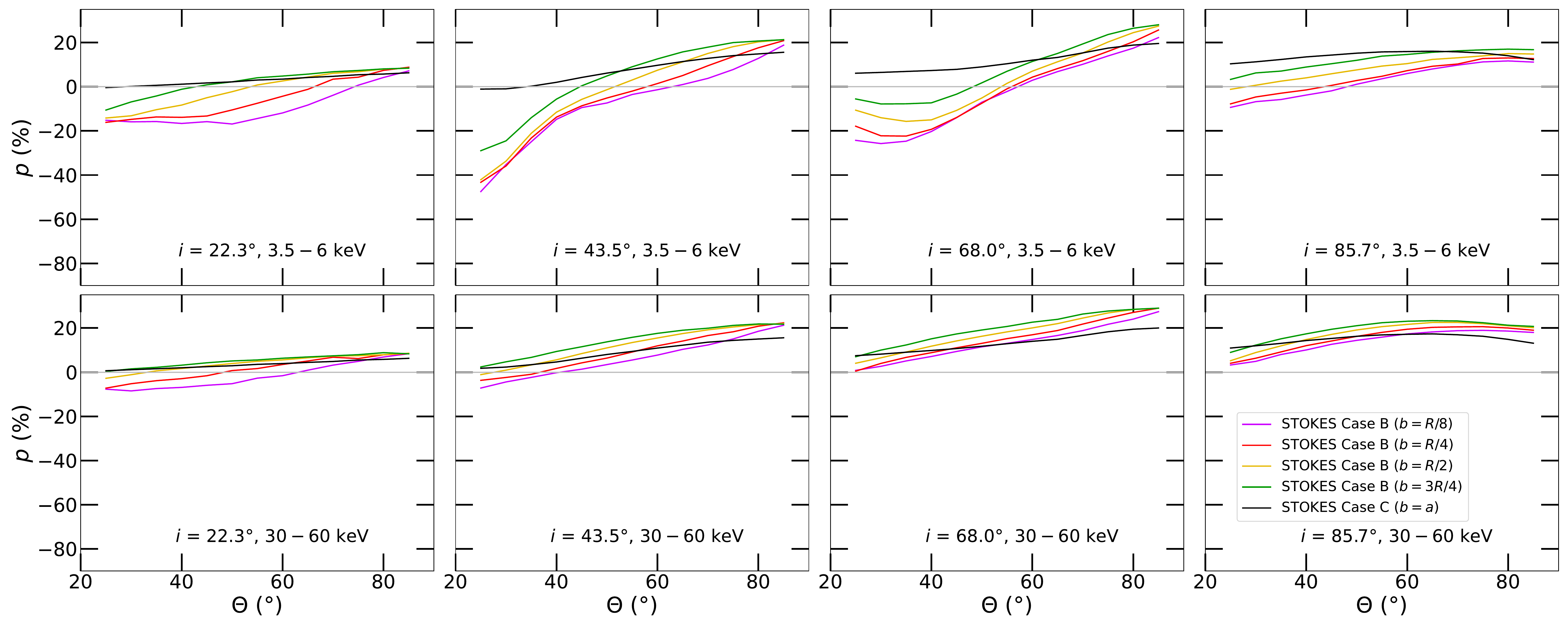}
	\caption{The same as in Figure \ref{BvsC_pmue24}, but energy-averaged polarization degree, $p$, versus half-opening angle $\Theta$ is shown for $i = 22.3\degr$, $i = 43.5\degr$, $i = 68.0\degr$, and $i = 85.7\degr$ (left to right). Different view of the previous results. For high energies the geometrical differences are minor.}
	\label{BvsC_ptheta24}
\end{figure*}

\begin{figure*}
	\includegraphics[width=2.\columnwidth]{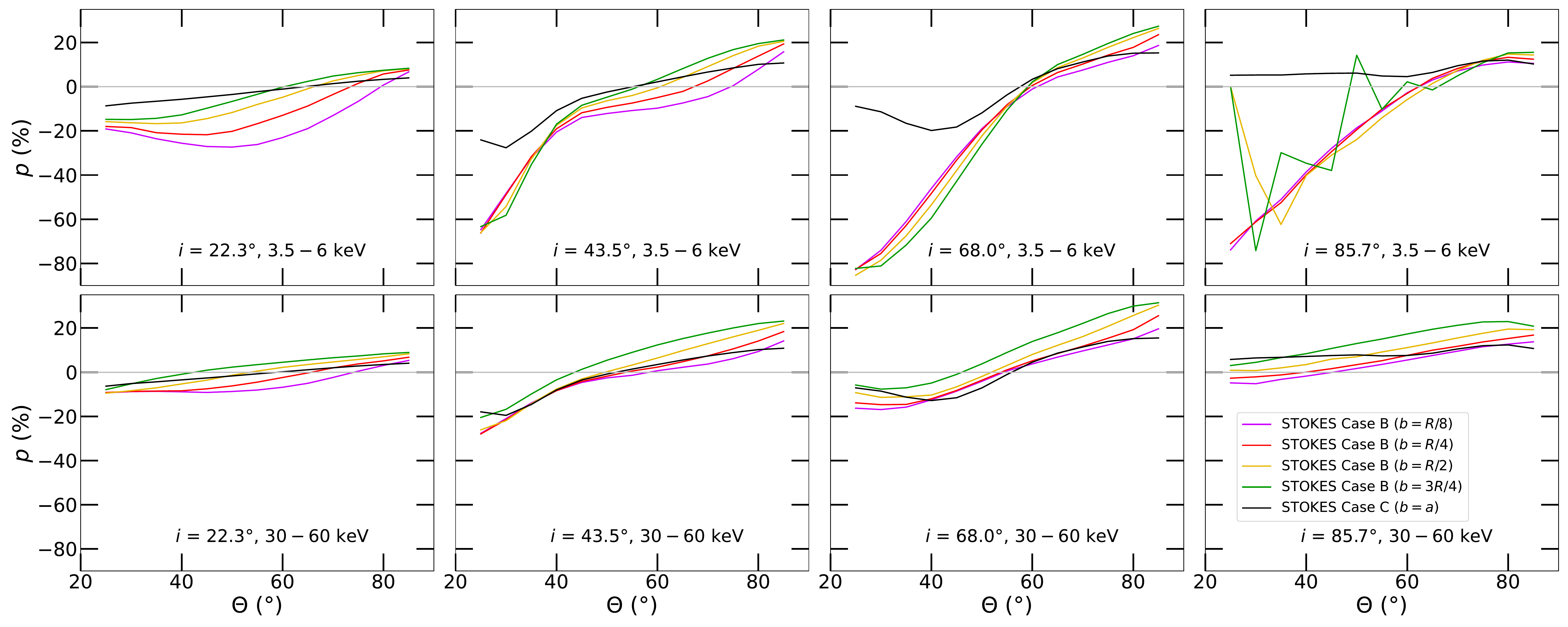}
	\caption{The same as in Figure \ref{BvsC_ptheta24}, but for $N_\mathrm{H} = 10^{25} \textrm{ cm}^{-2}$. Different view of the previous results. The high obscuration presented here causes dominant perpendicular polarization arising from scattering on the furthest side of the torus from observer's point of view, especially for low half-opening angles more prolate elliptical tori, if inclinations are moderate or low with respect to the axis of symmetry.}
	\label{BvsC_ptheta25}
\end{figure*}

We will now study the effect of changing eccentricities of elliptical tori only, fixing the convexity sign of the illuminated inner side of the obscurer (with respect to the source of emission). We compare in the same style in Figures \ref{BvsC_pmue24}-\ref{BvsC_ptheta25} the elliptical Case B for various $b$ values relative to $R$ and the circular Case C (with $a = b$).  Because the distance $r_\textrm{in}$ of the closest toroidal point to the center is fixed for all cases shown, and because we vary $R$ denoting the distance of the center and the elliptical center in the meridional plane, we track more precisely the shape difference between a compact torus and a geometrically extended torus. Due to a different computation of $a$ between Cases B and C (see Equations \ref{size_trafo} and \ref{opening_angle}), the Case C does not always produce an intermediate result between the shown Cases B. The more prolate and equatorially compact the torus is, the more perpendicular and higher polarization we expect. The more oblate the structure is, i.e. equatorially elongated, the lower perpendicular and higher parallel polarization we observe for the aforementioned reasons. In this sense, the oblate Case B with $b=3R/4$ is the most similar to the wedge-shaped torus results, providing similar high column density conditions for light rays seen under inclinations near the half-opening angle. For such cases the exact convex or concave shape of the inner-most wall plays little importance. The dependency on eccentricity is reversed only for low half-opening angles, in particular at soft energies, high densities and high inclinations, when the absorption effects cause high perpendicular polarization for more oblate tori under extreme obscurations (with low numerical noise seen e.g. in top-middle panel of Figure \ref{BvsC_pmue25} and in comparison with Case A in dashed green curve of bottom-middle panel of Figure \ref{AvsC_pmue}). This is also the example of configuration space corner where the reflection on the inner walls matters mostly for the net result and the curvature sign (whether convex or concave with respect to the center) of the inner reflecting surface affects the result. Direct comparisons with other models, such as in \cite{Ratheesh2021, Ursini2023, Veledina2023} are beyond the scope of this paper, but we anticipate that, if anywhere, the conical shape adopted therein with essentially no inner wall would produce different results than the elliptical tori presented in our study (and even more the wedge-shaped tori presented here) for this part of configuration space most clearly. Section \ref{cygx3_results} and Appendix \ref{additional} provide a different way of plotting the typical geometrical dependencies, which is perhaps more comparable to the modelling results presented in the other publications.

Figures \ref{BvsC_pmue24} and \ref{BvsC_pmue25}, representing the inclination dependency for lower and higher density cases, are displayed for high ionization. See Figures \ref{BvsC_pmue24_low_io} and \ref{BvsC_pmue25_low_io} for the same figures with low ionization of the scattering media. Although more numerical noise is present, because more radiation is absorbed for the same number of simulated photons, we see qualitatively the same sensitivity of polarization to changing scattering geometry. The results are quantitatively similar for low densities, but when we simulate high densities, we see smaller dispersion in polarization with respect to eccentricity in the parallel oriented regimes. This is because when low ionization is present, the absorption on neutral atoms under these densities is large enough compared to pure reflection on free electrons, even for the most prolate cases. Then the furthermost side of the torus towards the observer is not able to produce enough scattered radiation to depolarize the parallelly polarized component from closer visible regions towards the observer. To view the ionization effects more directly, we show the same polarization results versus inclination and half-opening angle on Figures \ref{CvsC_pmue} and \ref{CvsC_ptheta} for 3 different ionization levels in one plot for the Case C only.

\begin{figure*}
	\includegraphics[width=2.\columnwidth]{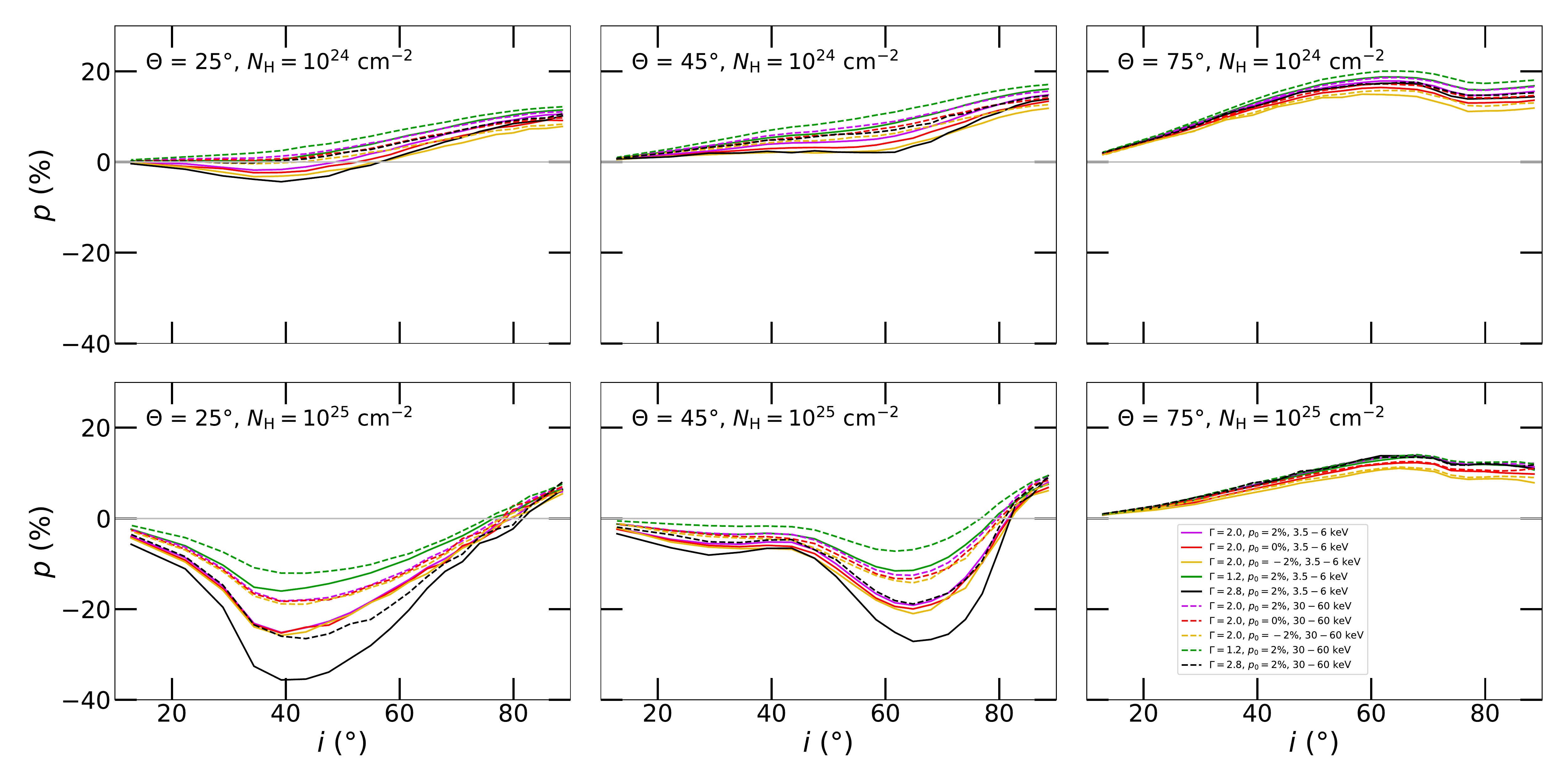}
	\caption{Case C only for different cases of primary radiation: energy-averaged polarization degree, $p$, versus inclination $i$ for $\Theta = 25\degr$ (left), $\Theta = 45\degr$ (middle), and $\Theta = 75\degr$ (right). We show the results integrated in 3.5--6 keV (solid) and 30--60 keV (dashed) and for $N_\mathrm{H} = 10^{24} \textrm{ cm}^{-2}$ and $\tau_\textrm{e} = 0.7$ (top) and $N_\mathrm{H} = 10^{25} \textrm{ cm}^{-2}$ and $\tau_\textrm{e} = 7$ (bottom). The magenta curves correspond to the 2\% parallelly polarized primary with $\Gamma = 2.0$. The red and yellow correspond to the unpolarized and 2\% perpendicularly polarized primary with $\Gamma = 2.0$, respectively. The green and black curves correspond to the 2\% parallelly polarized primary with $\Gamma = 1.2$ and $\Gamma = 2.8$, respectively. The results are shown for \textit{high} ionization, although other ionization levels show similar dispersion in polarization with respect to the changing primary input. This dispersion is different with respect to the chosen geometry, while the displayed Case C represents an intermediate result from all tested cases. The spectral index of incident emission drives the emergent polarization more strongly than the realistic cases of primary polarization tested.}
	\label{C_primary}
\end{figure*}

Lastly, let us discuss the effects of changing the incident radiation properties. Although we do not aim to study the imprints of extended and anisotropic emission sources (and the scattering regions in {\tt STOKES} are not even thermally radiating), Figure \ref{C_primary} shows the effects of changing $\Gamma$ between $1.2 \leq \Gamma \leq 2.8$ on the resulting polarization state for the Case C torus. We also include the effect of changing primary polarization between unpolarized and 2\% polarized central emission with a polarization vector oriented parallelly and perpendicularly to the axis of symmetry. Studying only the reprocessed light rays, we see the incident polarization state effect on the resulting polarization fraction of the order of a few degrees. The departure from the originally unpolarized emission is in the same direction, i.e. depolarizing or additionally polarizing depending on the resulting net polarization angle. The power-law index $\Gamma$ plays, however, a larger role in the resulting polarization. Even though we omitted the energy bands with emission lines, whose equivalent widths are also $\Gamma$-dependent, the relative contribution between the soft and hard X-rays is important for the energy-dependent absorption effects. For low $\Gamma$, we rather see parallel oriented polarization, originating on the left and right sides from the observer's perspective, because photo-electric opacities that are in general larger at soft X-ray energies allow in this case more photons passing through the torus edges. Therefore the polarized flux contribution for perpendicular polarization originating in highly reduced geometry for high obscuration is weaker. The opposite is at work for large $\Gamma$. The difference can be even as large as $\sim 20$\% in polarization fraction in 3.5--6 keV for high densities, moderate inclinations and low half-opening angles (see Section \ref{cygx3_results} for the discussion on degeneracy of the parameter space with respect to a single 2--8 keV polarimetric measurement). We note that Figure \ref{C_primary} shows the result for high ionization only, but when inspecting other levels of partial ionization we observe similar dispersion in polarization with respect to the changing primary input. However, the dispersion is different with respect to the chosen geometry. The displayed circular Case C represents an intermediate result from all tested cases. The more prolate elliptical tori are the most sensitive to $\Gamma$ (up to $\sim 30$\% difference in polarization in the most sensitive configurations in 3.5--6 keV), while for the more oblate elliptical tori and the wedge-shaped tori the differences are the smallest (up to $\sim 10$\% in 3.5--6 keV). Regarding application to XRBs, a question arises on how the presence of thermal peak at soft X-rays originating from inner accretion would affect the results on top of the tested power-law. Assuming that the region of emission would still have negligible size and location in the center, we can draw upon the above conclusions. The IXPE observations of 4U1630-47 \citep{Cavero2023} reveal no significant change in polarization, despite a spectral state change. But even if the thermal component was producing different polarization, the changing flux contribution at soft X-rays due to thermal radiation would be of much higher importance, as our results for varying $\Gamma$ with respect to changing primary polarization indicate. However, any more detailed quantitative estimate would require new simulations.

The effects of partial transparency, changing ionization and the polarization and slope of the primary power-law on the resulting polarization are directly related to the X-ray band. They shall be carefully X-ray spectroscopically analyzed for X-ray spectro-polarimetric observational campaigns in order to reduce the shown polarization degeneracies. The default geometrical parameters of the obscurer, such as $\Theta$, $i$ and orientation of the system axis of symmetry, are in addition directly shared for the same medium examined in other wavelengths. In particular, IR, optical and UV polarimetry of AGNs is able constrain the torus geometrical properties, alongside hints on the geometry of the broad line regions or polar scatterers with respect to the inner sub-parsec accretion engine \citep[e.g.][]{Antonucci1985, Oliva1998, Smith2004, Gaskell2012, Marin2018d, Hutsemekers2019, Marin2020, Stalevski2023}. IR and optical polarimetry alongside radio and optical imaging of accreting compact objects allows to constrain large-scale jet orientation, which can then be linked to the orientation of the accretion disc, knowing how the particle acceleration originates \citep[e.g.][]{Unger1987,Jones1994,Marti2000,Stirling2001,MillerJones2004,Cooke2008, Krawczynski2022, Kravtsov2023}.

\section{Application to X-ray binary system Cygnus X-3}\label{cygx3_results}

Throughout this section (in contrast to Section \ref{equatorial}), we register and display all unabsorbed photons, including the unscattered emission, if not stated otherwise. The inclusion of zero-scattered photons is important for the type-1 viewing angles, as the direct flux contribution ultimately dilutes the observed polarization. We set $R = 0.5$ constant in this section and let $r_\textrm{in}$ vary, referring to the results in Section \ref{equatorial} where $R$ was a variable and $r_\textrm{in}$ fixed. Again, the true units are unimportant for the presented polarization studies. We also provide an alternative way of displaying the results, similar to \cite{Veledina2023}, so that the MC modelling is easily compared to the analytical modelling therein, or to the equatorial obscuring MC simulations from \cite{Ghisellini1994}, displayed in \cite{Ursini2023} (see their Figure 6) for the X-ray polarimetric analysis of the case of obscured Circinus Galaxy. Finally, as mentioned in Section \ref{methods}, we do not account for neutral hydrogen to simulate the hydrogen-poor case of Cygnus X-3 accreting stellar-mass black hole and to trace the chemical composition effects. The $N_\textrm{He}$ and $\tau_\textrm{e}$ are scaled accordingly to represent similar obscuration and partial ionization levels to the previous sections. We will first show the results integrated in the 2--8 keV band, which is the full energy range of the IXPE mission, and directly compare the models to the latest Cygnus X-3 data. Later we will discuss the energy-dependence of polarization. We show the results only for $\Gamma = 2$ and unpolarized primary, but we performed the same set of simulations also for $\Gamma = 3$ and for $4\%$ parallelly and perpendicularly polarized primary radiation, which we discuss in words.

\begin{figure*}
	\includegraphics[width=2.\columnwidth]{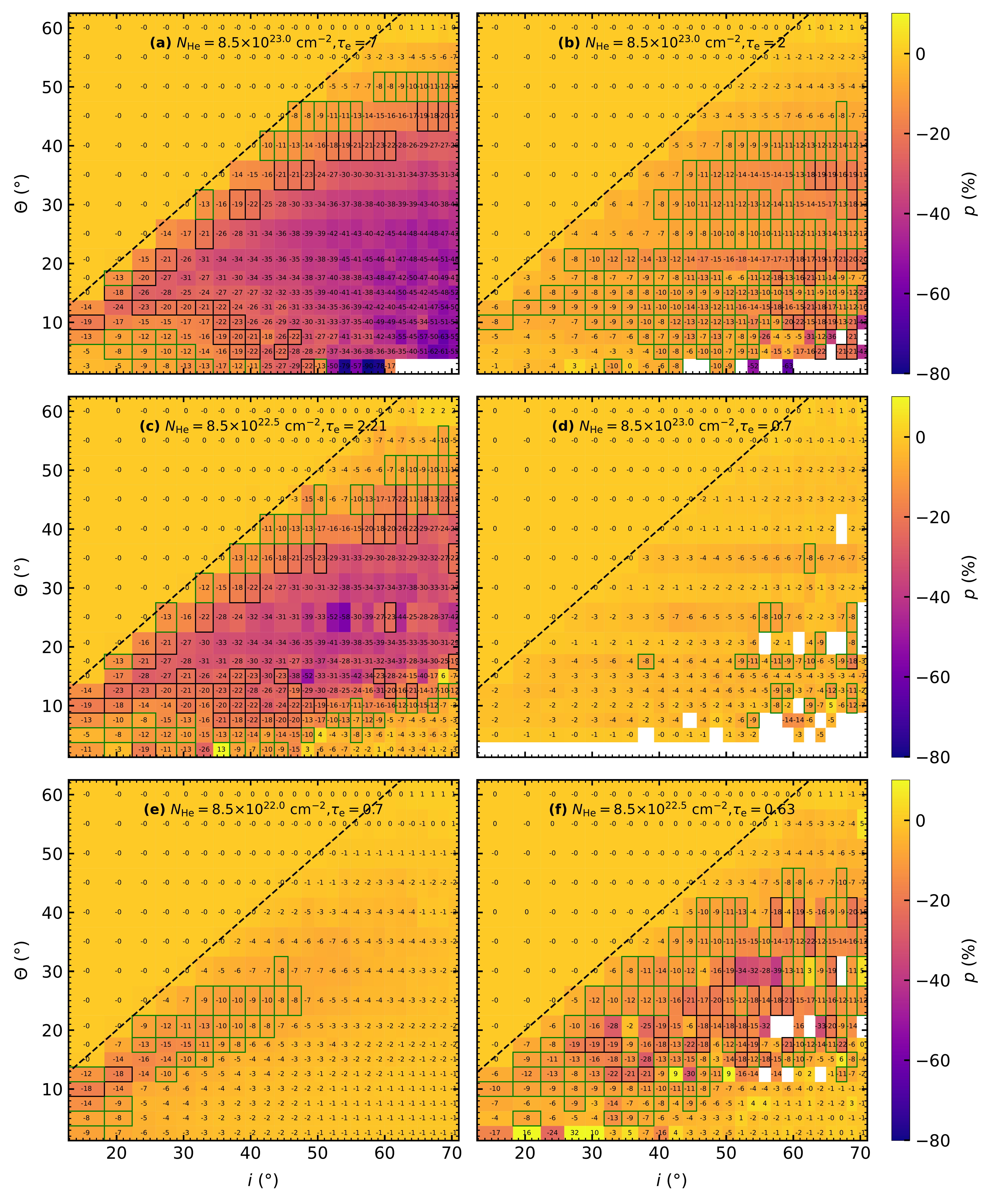}
	\caption{In the color code we show the {\tt STOKES} model polarization degree, $p$, integrated in 2--8 keV for the Case B elliptical equatorial obscurer with $b = R/4$ with respect to various inclinations $i$ and half-opening angles $\Theta$. All photons are registered and the black dashed lines marks the grazing inclination angles. The black and green rectangles represent the simulation results roughly matching the main and ToO IXPE observations, respectively, of an X-ray binary system Cygnus X-3, extensively described in \citet{Veledina2023}. We show the case of unpolarized primary radiation with $\Gamma = 2$. The left column shows the results for \textit{high} ionizations, how they differ when lowering the column densities, i.e. from top to bottom we show (a) $N_\textrm{He} = 8.5 \times 10^{23.0} \ \textrm{cm}^{-2}$ and $\tau_\textrm{e} = 7$, (c) $N_\textrm{He} = 8.5 \times 10^{22.5} \ \textrm{cm}^{-2}$ and $\tau_\textrm{e} = 2.21$, (e) $N_\textrm{He} = 8.5 \times 10^{22.0} \ \textrm{cm}^{-2}$ and $\tau_\textrm{e} = 0.7$. In top right and center right we show the highest density case for \textit{lower} ionization levels, i.e. (b) $N_\textrm{He} = 8.5 \times 10^{23.0} \ \textrm{cm}^{-2}$ and $\tau_\textrm{e} = 2$ and (d) $N_\textrm{He} = 8.5 \times 10^{23.0} \ \textrm{cm}^{-2}$ and $\tau_\textrm{e} = 0.7$, respectively. The (f) bottom right panel shows the moderate density case for \textit{lower} ionization, i.e. $N_\textrm{He} = 8.5 \times 10^{22.5} \ \textrm{cm}^{-2}$ and $\tau_\textrm{e} = 0.63$. The way of displaying results is identical to the MC simulation in Figure 6 of \citet{Ursini2023} and to the analytical modelling presented in \citet{Veledina2023}.}
	\label{all_torus_0.25}
\end{figure*}

Figure \ref{all_torus_0.25} shows the resulting polarization in $\{ i, \Theta \}$ plane in the 2--8 keV band for an elliptical torus, i.e. Case B, with $b = R/4$ and various column densities and ionization levels. The dashed black line marks the grazing inclination angles. The black rectangles represent the main IXPE observation in 2--8 keV from October 2022, i.e. $p = 21 \pm 3$ \% with simulation uncertainty, and the green rectangles represent the additional Target of Opportunity (ToO) observation from December 2022, i.e. $p = 10 \pm 3$ \% with simulation uncertainty, interpreted in \cite{Veledina2023} as depolarization due to higher transparency of the equatorially obscuring funnel. The empty bins (in white), that are typically at high inclinations and low half-opening angles, did not converge within reasonable computational time of the MC simulations ($\sim 2$ months at the available institutional cluster capacities). Despite the modelling changes with respect to the previous sections, and despite the inclusion of spectral lines in the main geometrical dependencies shown in the entire 2--8 keV band, we obtained comparable polarization prediction for the type-2 viewing angles, i.e. similar linear polarization degree and angle and their dependencies on inclination and half-opening angle of the obscurer. For almost all tested type-2 viewing angles, we obtained perpendicular polarization due to high obscuration and dominant reflection off the inner walls of the furthermost side of the reflecting funnel.

In Figures \ref{all_torus_0.125}--\ref{all_flared} we show the same panels for elliptical Case B tori with $b = R/8$ and $b = 3R/4$ and the wedge-shaped Case A. The effect of constant $R$ and varying $r_\textrm{in}$ with respect to different elliptical eccentricities is similar to that of Figure 7 in \cite{Goosmann2007}, studying the compactness of the torus in optical and UV radiation. We show that also in the X-rays the large tori scatter fewer photons in the direction of the observer, because of the geometrically extended surface area that reflects them, resulting in less reduced geometry of scattering and lower polarization. In this way, we see larger differences between the geometrical archetypes. The two more prolate elliptical tori show rather similar behavior in resulting polarization (up to about $\sim 10\degr$ error in $i$ and $\Theta$ for the full-band IXPE results between $b = R/8$ and $b = R/4$), but once we switch to the most oblate cases of $b = 3R/4$, the results are similar to those of the wedge-shaped torus. The reason is simple: for the latter the line-of-sight column densities are large shortly below the grazing inclination angle, where we see very high polarization degree already. Ultimately, for the Case A the line-of-sight column density is a sharp step function of inclination from zero and does not change further with inclination for any type-2 observer.

For the type-1 viewing angles we obtain typically very low polarization -- following the incident polarization state. Compared to the results in the previous sections, this is due to the inclusion of the zero-scattered photons that are dominating in the total flux when the observer can see directly into the accreting nucleus. We conclude that for type-1 observers, reprocessing changes the original polarization by $\sim 1$\% in polarization fraction in the 2--8 keV band, depending on the equatorial environment, which may still be relevant for the interpretation of the IXPE observations of type-1 AGNs and other unobscured accreting compact objects. On the other hand the zero-scattered photons play negligible role in resulting mid X-ray polarization for the Compton-thick sources.

Observing the polarization change with column density, we confirm the possible interpretation of the depolarization by $\sim 10\%$ in Cygnus X-3 due to scatterings in larger volumes of lower column densities, thus diluting the preferred scattering-plane orientation inside the system. When testing even lower densities than those displayed, we obtained nearly unpolarized emission. Similarly to density, lowering the ionization level can also completely depolarize the output for type-2 viewing angles due to effectively higher symmetry and absorption of otherwise highly collimated and highly polarized photons. In Figures \ref{all_torus_0.25} and \ref{all_torus_0.125}--\ref{all_flared} we also see a different manifestation of depolarization with lowering density and ionization level with inclination per one $\Theta$. For type-2 inclinations just below the grazing angle, we see the higher depolarization due to changing ionization via lowering the homogenous free electron density. This is because the transparency of surface layers of the toroidal structures matters for changing ionization. On the other hand when lowering the uniform density of ions while keeping the ionization level roughly equivalent, we first see depolarization at moderate and higher type-2 inclinations than those just below the half-opening angle. This is because the change occurs in the entire scattering location. Its shape, elongation and relative size are important when increasing the volume for reprocessing consistently. For future X-ray polarimetric observations, the depolarization due to ionization level is perhaps more likely to be contrained from simultaneous X-ray spectroscopy than Compton thickness.

The inclination, half-opening angle, density and ionization level seem to play much larger role in resulting polarization than any other tested parameters. For all tested geometries, the $\Gamma = 3$ cases and/or the $4\%$ polarized cases lead to an error in derived $i$ or $\Theta$ of $\sim 10\degr$ for type-2 viewing angles. This is negligible compared to the arbitrary polarization (up to tens of \%) that is given by the main four parameters that have to be in prior constrained from X-ray spectroscopy, if one desires to interpret the IXPE polarization data in the full energy band. In this way, we show the level of degeneracy in parameter constraint through a single 2--8 keV polarization data of a particular source. The actual chemical composition seems to be of low overall importance unless we have ambitions to study contribution of individual spectral lines to polarization, and ionization edges in the photo-electric opacities in detail, which is beyond the energy resolution adopted, and sensitivity of current and near-future X-ray polarimeters, and scope of this study. 

The contour maps of solutions in the $\{ i, \Theta \}$ space show two branches of solutions, both compatible with the single-scattering analytical model presented in \cite{Veledina2023} and the upper branch being compatible with the modelling result shown in \cite{Ursini2023}, using the \cite{Ghisellini1994} model, which did not provide low enough half-opening angles to see the lower branch of solutions. Both of these models assumed a conical shape of the torus, which means that also the true geometrical shape is of lower importance for the resulting polarization fraction than the aforementioned parameters. However, a detailed one-to-one comparison is required for further analysis of the different configurations and methods.

\begin{figure*}
	\includegraphics[width=2.1\columnwidth]{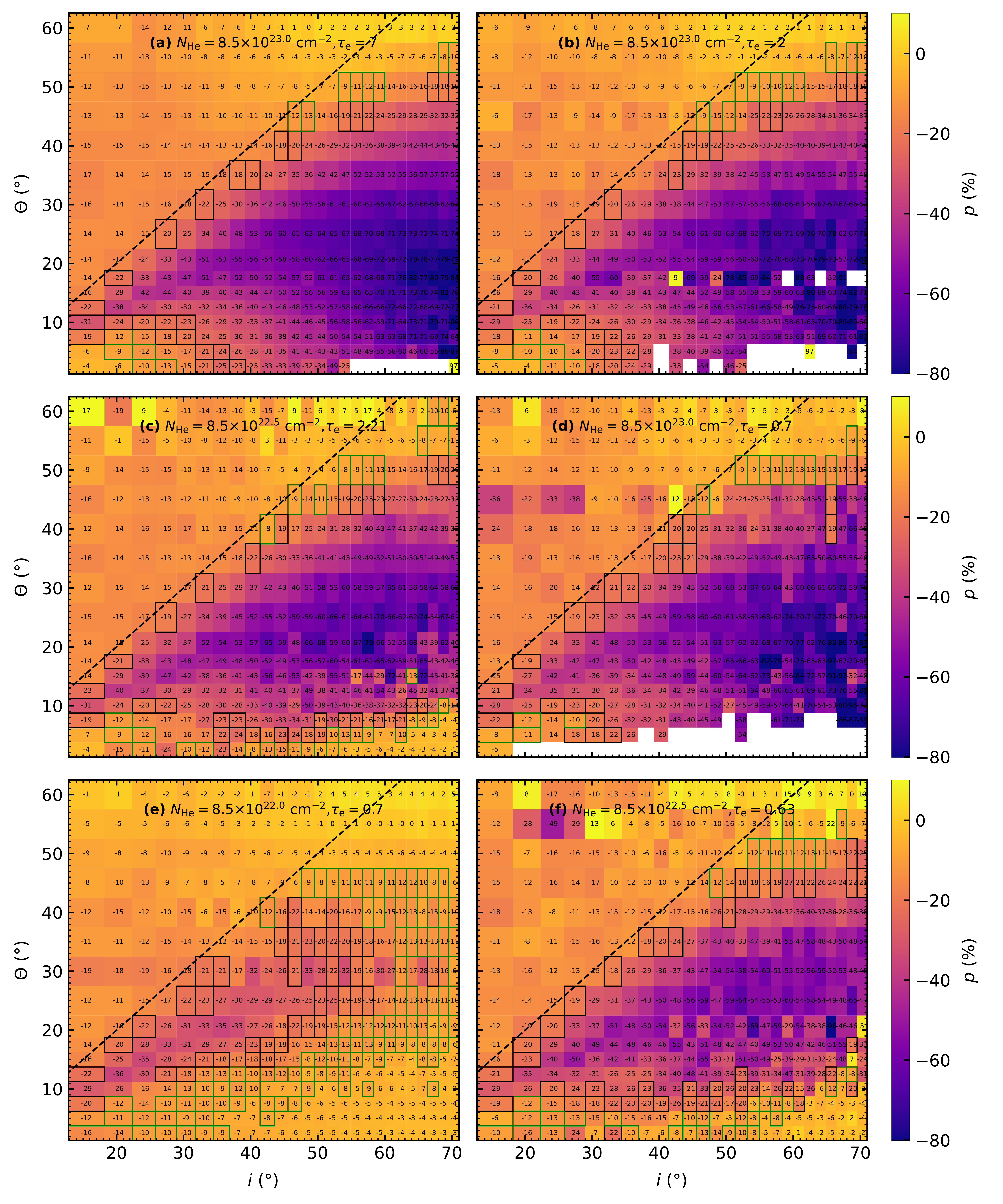}
	\caption{The same as in Figure \ref{all_torus_0.25}, but for registered single-scattered photons only. Compared to the simulations with all photons, the polarization still contains two branches of solutions in the \{$i$,$\Theta$\} space for type-2 viewing angles, but the average obtained polarization is higher and the dependency on free electron density is negligible.}
	\label{single_torus_0.25}
\end{figure*}

{\tt STOKES} is unique in a sense that it can provide the effects of partial ionization and partial transparency. We approximate the results of the other two models best in the high ionization and high density cases. The Figure 6 in \cite{Ursini2023} follows the \cite{Ghisellini1994} model with neutral matter and thus provides in general perhaps more dissimilar predictions from the {\tt STOKES} code than the analytical model presented in \cite{Veledina2023}, which targets single-scattering on electrons. Moreover, the latter model was directly applied to the Cygnus X-3 example discussed in this section and we plan to focus on the dusty obscurer in the Circinus Galaxy with {\tt STOKES} in a separate publication. The analytical model presented in \cite{Veledina2023} is representing the single-scattering inner-wall reflection, which we tried to approach with {\tt STOKES} in a separate round of simulations where we detected only the single-scattered photons. We plotted the results in the same way in Figure \ref{single_torus_0.25}, for the same geometry of scattering as in Figure \ref{all_torus_0.25}, but similar conclusions apply when comparing the single-scattering results to the full simulations in other geometries. Clearly, we are much closer to the analytical model in the type-2 viewing angles than when all radiation is taken into account. We note that there is still a principal difference in between the two models that in {\tt STOKES} even in this single-scattering regime we work with absorption of photons.

\begin{figure*}
	\includegraphics[width=2.15\columnwidth]{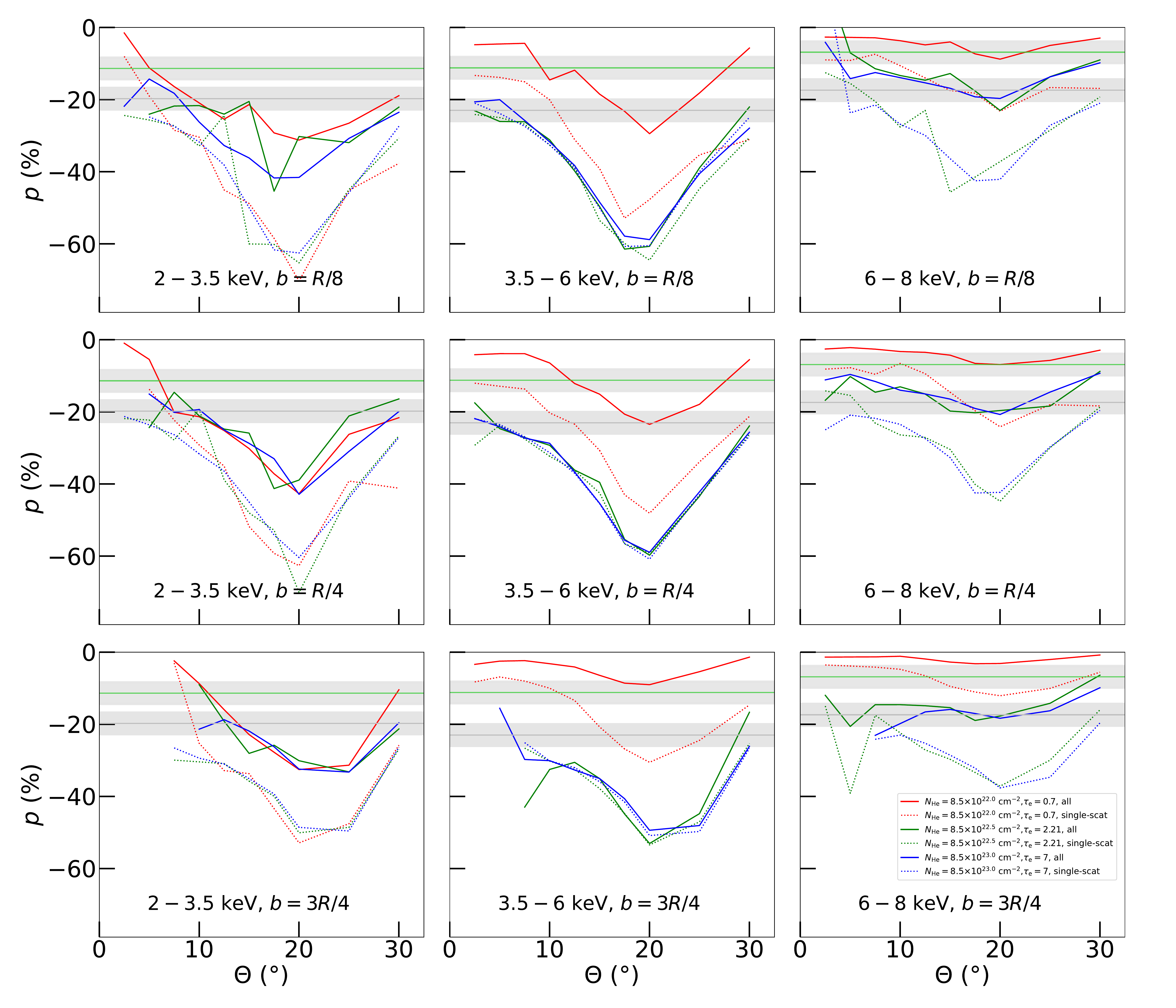}
	\caption{The {\tt STOKES} model polarization degree, $p$, averaged in 2--3.5 keV (left), 3.5--6 keV (middle) and 6--8 keV (right), versus half-opening angle $\Theta$ (in the obscured type-2 configurations only) for a selected $i = 30\degr$ inclination angle, corresponding to the orbital inclination in Cygnus X-3 \citet{Vilhu2009, Antokhin2022}. The corresponding energy-averaged polarization data from the main and ToO observations are marked with gray and green horizontal lines with shadowed areas corresponding to simulation uncertainties. The top, middle and bottom panels correspond to the Case B simulations of elliptical tori with eccentrities given by $b = R/8$, $b = R/4$ and $b = 3R/4$, respectively. We show the results for all registered photons (solid lines) and only those photons that scattered once (dotted lines). The different colors correspond to various column densities for \textit{high} ionization: $N_\textrm{He} = 8.5 \times 10^{23.0} \ \textrm{cm}^{-2}$ and $\tau_\textrm{e} = 7$ (blue), $N_\textrm{He} = 8.5 \times 10^{22.5} \ \textrm{cm}^{-2}$ and $\tau_\textrm{e} = 2.21$ (green), $N_\textrm{He} = 8.5 \times 10^{22.0} \ \textrm{cm}^{-2}$ and $\tau_\textrm{e} = 0.7$ (red). Apart from low densities, the total result is well represented by the single-scattering simulation in 3.5--6 keV band. The density affects absorption, which is energy dependent and results in lower polarization at higher continuum energies.}
	\label{slices_1_torus_0.25}
\end{figure*}

\begin{figure*}
	\includegraphics[width=2.15\columnwidth]{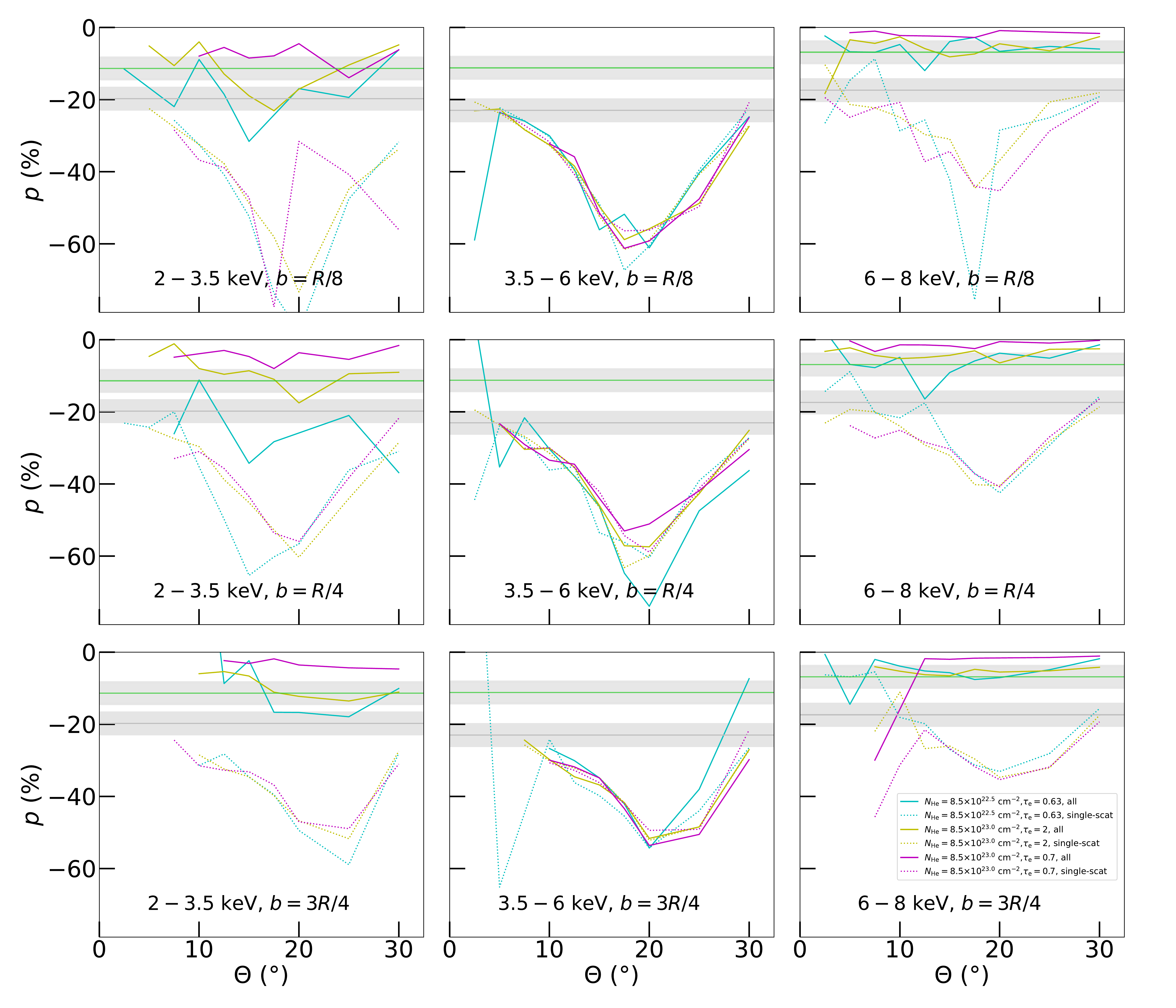}
	\caption{The same as in Figure \ref{slices_1_torus_0.25}, but for \textit{lower} ionization levels. For the highest density case $N_\textrm{He} = 8.5 \times 10^{23.0} \ \textrm{cm}^{-2}$ we show $\tau_\textrm{e} = 2$ (yellow) and $\tau_\textrm{e} = 0.7$ (magenta). For the moderate density case $N_\textrm{He} = 8.5 \times 10^{22.5} \ \textrm{cm}^{-2}$ we show $\tau_\textrm{e} = 0.63$ (cyan). The amount of ionization affects the presence of spectral lines, which depolarize at the lower and upper energy band shown (left and right columns, respectively).}
	\label{slices_2_torus_0.25}
\end{figure*}

The model degeneracy for the type-2 viewing angles with respect to the 2--8 keV data is further reduced when analyzing the polarization energy-dependence within the 2--8 keV range. This is in practice more achievable for Galactic sources than for AGNs, owing to the larger average count rates. The wedge-shaped model shown on Figure \ref{all_flared} shows too high obscuration for the high line-of-sight column densities just below the grazing inclination angles. Thus, we do not have enough MC photon statistics to study the energy-dependence with the performed MC runs even in theory -- apart from noticable depolarization in 6--8 keV compared to lower energies, which occurs clearly due to the dominant iron emission line present in all studied spectra. With the performed simulations, detailed energy-dependence of polarization is however possible to examine for the partially transparent elliptical tori when integrated in the 2--3.5 keV, 3.5--6 keV and 6--8 keV bands. This is shown on Figures \ref{slices_1_torus_0.25} and \ref{slices_2_torus_0.25}. We display the energy-dependent results for the approximate orbital inclination of Cygnus X-3, $i = 30\degr$ \citep{Vilhu2009, Antokhin2022} (which can still be different with respect to the true average accretion disc orientation) and for lower half-opening angles and various combinations of densities and ionizations. The numerical noise is too high to provide acceptable geometry and composition constraints for both IXPE Cygnus X-3 observations represented by the gray and green horizontal lines, which should match within one branch of solutions.

First of all, we prove that the single-scattering analytical model discussed in \cite{Veledina2023} well represents the general {\tt STOKES} model in 3.5--6 keV band where no prominent spectral lines are present. Apart from spectral lines, the energy-dependent bound-free and free-free absorption also impact the energy-dependence of polarization, which is inherently related to the geometrical parameters of the system. The {\tt STOKES} single-scattering approximation (compared here directly to the full {\tt STOKES} model) is valid in the 3.5--6 keV in all cases apart from the lowest column densities of neutral species (red lines in Figure \ref{slices_1_torus_0.25}). In this case the depolarization in continuum energies is too high due to enlarged angular distribution of scatterings. Although the conditions of line formation can be complex with respect to the level of depolarization caused on the underlying continuum radiation, our simulations show that at 2--3.5 keV and 6--8 keV the fluorescent spectral lines may noticeably depolarize the observed spectrum compared to the values of polarization at 3.5--6 keV band. The stronger depolarization is more likely for lower ionization fractions, as expected from the line formation. The depolarization through changing the $N_\textrm{He}$ density is sensitive with respect to the particular Compton-thickness energy transition, which occurs in or near the 2--8 keV band due to photo-electric opacities, being comparable to Compton opacities. We observe smaller decrease of polarization degree from 3.5-6 keV to 2-3.5 keV (increase actually in some low $N_\textrm{He}$ density cases) when compared to the depolarization from 3.5-6 keV to 6--8 keV, because the obscuring medium is indeed more transparent for the central source photons at higher energies, which causes monotonous depolarization with increasing energy (see Figure \ref{energy_dependent}), unlike the spectral lines effect. This would explain the leveling of polarization degree at 2--6 keV with energy in the ToO observation of Cygnus X-3 with IXPE \citep{Veledina2023}, if the source was caught in obscuring outflows with lower total column density during the second observation period. However, our models are largerly simplified due to the uniform density assumption. The effects of obscurers with non-uniform density on polarization are left for future studies.

\section{Conclusions}\label{conclusions}

We compared the X-ray polarization properties in the 1--100 keV band of homogenous static equatorial obscuring media around accreting compact objects. We performed simulations of partially ionized and partially transparent 3D toroidal structures with the {\tt STOKES} code \citep{Goosmann2007,Marin2012,Marin2015,Marin2018}, which is an MC simulation suitable for studying the effects of scattering and absorption. Placing the source of low (up to a few \%) polarized source of isotropic power-law emission in the center of the axially symmetric equatorial obscurer, we defined distinct geometries for the reprocessing studies: the wedge-shaped torus and elliptical tori with various eccentricities. When avoiding the observed polarization state of spectral lines by selecting the 3.5--6 keV and 30--60 keV bands, we found large dependence of polarization degree on inclination, half-opening angle, density of the scatterer and ionization level. We found relatively minor impact (up to a few \%) of changing primary polarization state by 2\% from unpolarized emission and moderate impact of changing primary power-law index $\Gamma$ by 0.8 from 2.0 (low tens of \% in very specific geometries). Overall, the diversity in resulting polarization is striking. The polarization fraction of \textit{reprocessed-only} radiation can vary from completely unpolarized results up to high tens of \%, with large degeneracies with respect to the geometries examined and the density and ionization effects. The resulting polarization angle can be either perpendicular or parallel with respect to the main axis of symmetry, as we do not take into account any general-relativistic effects. The former scenario is favoured for low half-opening angles, high column densities, high ionization levels and low inclinations, causing the dominant single-scattering plane to be meridional. The opposite configurations will more likely result in a parallelly oriented net polarization angle with equatorial dominant single-scattering plane, while the intermediate cases depend on subtle details of relative size, shape and composition of the obscurer. The X-ray polarization fraction typically reaches a maximum for moderate inclinations per given half-opening angle, effectively providing 2 solutions as a diagnostic tool for energy-integrated observations. The modelling results are roughly consistent with other codes in the literature \citep{Ghisellini1994, Ratheesh2021, Ursini2023, Veledina2023} and {\tt STOKES} code computations performed for more specific configurations \citep{Goosmann2011}. We obtained similar geometry effects of the equatorial obscurers on polarization studied in the optical and UV wavelengths \citep{Kartje1995, Goosmann2007}, confirming the fundamental geometrical explanations provided therein. This is because the physical mechanisms behind generation of polarization are alike: absorption that selects the dominant reprocessing geometry with respect to the observer and scatterings that additionally modify polarization of individual photons. The main differences are caused by individual spectral lines and relative contribution of absorption and scattering opacities, which is energy dependent.

In addition, we provided a different perspective and usage example of the MC simulations by studying in detail the equatorial obscuration in the context of recent IXPE observation of the XRB Cygnus X-3 \citep{Veledina2023} with severe hydrogen depletion. Focusing on the IXPE operational band 2--8 keV, we showed maps of predicted polarization state with respect to inclination, half-opening angle, different toroidal geometries, equatorial column densities and ionization levels for single-scattered photons and \textit{all photons} (including direct radiation). For constant density, ionization level and geometrical parameters, the actual chemical composition seems to have low effect, if one does not have ambitions to study the polarization signatures in spectral lines in detail. In the continuum 3.5--6 keV band we obtained a match between the moderate and high density full MC simulations with the single-scattering option and the analytical single-scattering results presented in \cite{Veledina2023}.
For type-2 viewing angles the primary state of emission is washed out up to $\sim 10\degr$ of error in inclination and half-opening angle, which is negligible compared to the effects of changing density and ionization. Although spectroscopy and polarimetry in other wavelengths can significantly help to lift the model degeneracies, X-ray polarimetry can do so already, if energy-dependent polarization is obtained with sufficient photon statistics in the 2--8 keV band, which is particularly sensitive to partial transparency effects and spectral lines. This was also illustrated on the recent IXPE observation of the obscured accreting stellar-mass black hole in Cygnus X-3 in comparison with the MC models within the limits of unavoidable numerical noise for reasonable computational times.

\section*{Acknowledgements}

The authors thank Alexandra Veledina for consulting the modelling of the Cygnus X-3 source, and the Strasbourg Astronomical Observatory for providing the necessary computational capacities. JP and MD also acknowledge the support from the Czech Science Foundation project GACR 21-06825X and the institutional support from the Astronomical Institute RVO:67985815.

\section*{Data Availability}

The {\tt STOKES} code version \textit{v2.07} is currently not publicly available and it will be shared upon reasonable request.



\bibliographystyle{mnras}
\bibliography{example} 

\begin{thebibliography}{}
\makeatletter
\relax
\def\mn@urlcharsother{\let\do\@makeother \do\$\do\&\do\#\do\^\do\_\do\%\do\~}
\def\mn@doi{\begingroup\mn@urlcharsother \@ifnextchar [ {\mn@doi@} {\mn@doi@[]}}
\def\mn@doi@[#1]#2{\def\@tempa{#1}\ifx\@tempa\@empty \href {http://dx.doi.org/#2} {doi:#2}\else \href {http://dx.doi.org/#2} {#1}\fi \endgroup}
\def\mn@eprint#1#2{\mn@eprint@#1:#2::\@nil}
\def\mn@eprint@arXiv#1{\href {http://arxiv.org/abs/#1} {{\tt arXiv:#1}}}
\def\mn@eprint@dblp#1{\href {http://dblp.uni-trier.de/rec/bibtex/#1.xml} {dblp:#1}}
\def\mn@eprint@#1:#2:#3:#4\@nil{\def\@tempa {#1}\def\@tempb {#2}\def\@tempc {#3}\ifx \@tempc \@empty \let \@tempc \@tempb \let \@tempb \@tempa \fi \ifx \@tempb \@empty \def\@tempb {arXiv}\fi \@ifundefined {mn@eprint@\@tempb}{\@tempb:\@tempc}{\expandafter \expandafter \csname mn@eprint@\@tempb\endcsname \expandafter{\@tempc}}}

\bibitem[\protect\citeauthoryear{Abarr et~al.,}{Abarr et~al.}{2021}]{Abarr2021}
Abarr Q.,  et~al., 2021, \mn@doi [Astroparticle Physics] {https://doi.org/10.1016/j.astropartphys.2020.102529}, 126, 102529

\bibitem[\protect\citeauthoryear{{Angel}, {Stockman}, {Woolf}, {Beaver}  \& {Martin}}{{Angel} et~al.}{1976}]{Angel1976}
{Angel} J.~R.~P.,  {Stockman} H.~S.,  {Woolf} N.~J.,  {Beaver} E.~A.,   {Martin} P.~G.,  1976, \mn@doi [\apjl] {10.1086/182120}, \href {https://ui.adsabs.harvard.edu/abs/1976ApJ...206L...5A} {206, L5}

\bibitem[\protect\citeauthoryear{{Antokhin}, {Cherepashchuk}, {Antokhina}  \& {Tatarnikov}}{{Antokhin} et~al.}{2022}]{Antokhin2022}
{Antokhin} I.~I.,  {Cherepashchuk} A.~M.,  {Antokhina} E.~A.,   {Tatarnikov} A.~M.,  2022, \mn@doi [\apj] {10.3847/1538-4357/ac4047}, \href {https://ui.adsabs.harvard.edu/abs/2022ApJ...926..123A} {926, 123}

\bibitem[\protect\citeauthoryear{{Antonucci}}{{Antonucci}}{1982}]{Antonucci1982}
{Antonucci} R.~R.~J.,  1982, \mn@doi [\nat] {10.1038/299605a0}, \href {https://ui.adsabs.harvard.edu/abs/1982Natur.299..605A} {299, 605}

\bibitem[\protect\citeauthoryear{{Antonucci}}{{Antonucci}}{1993}]{Antonucci1993}
{Antonucci} R.,  1993, \mn@doi [\araa] {10.1146/annurev.aa.31.090193.002353}, \href {https://ui.adsabs.harvard.edu/abs/1993ARA&A..31..473A} {31, 473}

\bibitem[\protect\citeauthoryear{{Antonucci} \& {Miller}}{{Antonucci} \& {Miller}}{1985}]{Antonucci1985}
{Antonucci} R.~R.~J.,  {Miller} J.~S.,  1985, \mn@doi [\apj] {10.1086/163559}, \href {https://ui.adsabs.harvard.edu/abs/1985ApJ...297..621A} {297, 621}

\bibitem[\protect\citeauthoryear{{Asplund}, {Grevesse}  \& {Sauval}}{{Asplund} et~al.}{2005}]{Asplund2005}
{Asplund} M.,  {Grevesse} N.,   {Sauval} A.~J.,  2005, in {Barnes} Thomas~G. I.,  {Bash} F.~N.,  eds,  Astronomical Society of the Pacific Conference Series Vol. 336, Cosmic Abundances as Records of Stellar Evolution and Nucleosynthesis. p.~25

\bibitem[\protect\citeauthoryear{Baes \& Camps}{Baes \& Camps}{2015}]{Baes2015}
Baes M.,  Camps P.,  2015, \mn@doi [Astronomy and Computing] {https://doi.org/10.1016/j.ascom.2015.05.006}, 12, 33

\bibitem[\protect\citeauthoryear{Baes, Verstappen, Looze, Fritz, Saftly, Pérez, Stalevski  \& Valcke}{Baes et~al.}{2011}]{Baes2011}
Baes M.,  Verstappen J.,  Looze I.~D.,  Fritz J.,  Saftly W.,  Pérez E.~V.,  Stalevski M.,   Valcke S.,  2011, \mn@doi [The Astrophysical Journal Supplement Series] {10.1088/0067-0049/196/2/22}, 196, 22

\bibitem[\protect\citeauthoryear{Beheshtipour, Krawczynski  \& Malzac}{Beheshtipour et~al.}{2017}]{Beheshtipour2017}
Beheshtipour B.,  Krawczynski H.,   Malzac J.,  2017, \mn@doi [ApJ] {10.3847/1538-4357/aa906a}, \href {https://ui.adsabs.harvard.edu/abs/2017ApJ...850...14B} {850, 14}

\bibitem[\protect\citeauthoryear{{Buchner}, {Brightman}, {Nandra}, {Nikutta}  \& {Bauer}}{{Buchner} et~al.}{2019}]{Buchner2019}
{Buchner} J.,  {Brightman} M.,  {Nandra} K.,  {Nikutta} R.,   {Bauer} F.~E.,  2019, \mn@doi [\aap] {10.1051/0004-6361/201834771}, \href {https://ui.adsabs.harvard.edu/abs/2019A&A...629A..16B} {629, A16}

\bibitem[\protect\citeauthoryear{Camps \& Baes}{Camps \& Baes}{2015}]{Camps2015}
Camps P.,  Baes M.,  2015, \mn@doi [Astronomy and Computing] {https://doi.org/10.1016/j.ascom.2014.10.004}, 9, 20

\bibitem[\protect\citeauthoryear{Camps \& Baes}{Camps \& Baes}{2020}]{Camps2020}
Camps P.,  Baes M.,  2020, \mn@doi [Astronomy and Computing] {https://doi.org/10.1016/j.ascom.2020.100381}, 31, 100381

\bibitem[\protect\citeauthoryear{{Cooke}, {Bland-Hawthorn}, {Sharp}  \& {Kuncic}}{{Cooke} et~al.}{2008}]{Cooke2008}
{Cooke} R.,  {Bland-Hawthorn} J.,  {Sharp} R.,   {Kuncic} Z.,  2008, \mn@doi [\apjl] {10.1086/593169}, \href {https://ui.adsabs.harvard.edu/abs/2008ApJ...687L..29C} {687, L29}

\bibitem[\protect\citeauthoryear{{Dibai} \& {Shakhovskoi}}{{Dibai} \& {Shakhovskoi}}{1966}]{Dibai1966}
{Dibai} E.~A.,  {Shakhovskoi} N.~M.,  1966, Astronomicheskij Tsirkulyar, \href {https://ui.adsabs.harvard.edu/abs/1966ATsir.375....1D} {375, 1}

\bibitem[\protect\citeauthoryear{{Dorodnitsyn} \& {Kallman}}{{Dorodnitsyn} \& {Kallman}}{2010}]{Dorodnitsyn2010}
{Dorodnitsyn} A.,  {Kallman} T.,  2010, \mn@doi [\apjl] {10.1088/2041-8205/711/2/L112}, \href {https://ui.adsabs.harvard.edu/abs/2010ApJ...711L.112D} {711, L112}

\bibitem[\protect\citeauthoryear{{Dorodnitsyn} \& {Kallman}}{{Dorodnitsyn} \& {Kallman}}{2011}]{Dorodnitsyn2011}
{Dorodnitsyn} A.,  {Kallman} T.,  2011, \mn@doi [\apss] {10.1007/s10509-011-0656-3}, \href {https://ui.adsabs.harvard.edu/abs/2011Ap&SS.336..245D} {336, 245}

\bibitem[\protect\citeauthoryear{{Doroshenko} et~al.,}{{Doroshenko} et~al.}{2023}]{Doroshenko2023}
{Doroshenko} V.,  et~al., 2023, \mn@doi [arXiv e-prints] {10.48550/arXiv.2306.02116}, \href {https://ui.adsabs.harvard.edu/abs/2023arXiv230602116D} {p. arXiv:2306.02116}

\bibitem[\protect\citeauthoryear{{Fender}, {Hanson}  \& {Pooley}}{{Fender} et~al.}{1999}]{Fender1999}
{Fender} R.~P.,  {Hanson} M.~M.,   {Pooley} G.~G.,  1999, \mn@doi [\mnras] {10.1046/j.1365-8711.1999.02726.x}, \href {https://ui.adsabs.harvard.edu/abs/1999MNRAS.308..473F} {308, 473}

\bibitem[\protect\citeauthoryear{{Fritz}, {Franceschini}  \& {Hatziminaoglou}}{{Fritz} et~al.}{2006}]{Fritz2006}
{Fritz} J.,  {Franceschini} A.,   {Hatziminaoglou} E.,  2006, \mn@doi [\mnras] {10.1111/j.1365-2966.2006.09866.x}, \href {https://ui.adsabs.harvard.edu/abs/2006MNRAS.366..767F} {366, 767}

\bibitem[\protect\citeauthoryear{{Furui}, {Fukazawa}, {Odaka}, {Kawaguchi}, {Ohno}  \& {Hayashi}}{{Furui} et~al.}{2016}]{Furui2016}
{Furui} S.,  {Fukazawa} Y.,  {Odaka} H.,  {Kawaguchi} T.,  {Ohno} M.,   {Hayashi} K.,  2016, \mn@doi [\apj] {10.3847/0004-637X/818/2/164}, \href {https://ui.adsabs.harvard.edu/abs/2016ApJ...818..164F} {818, 164}

\bibitem[\protect\citeauthoryear{{Gaskell}, {Goosmann}, {Merkulova}, {Shakhovskoy}  \& {Shoji}}{{Gaskell} et~al.}{2012}]{Gaskell2012}
{Gaskell} C.~M.,  {Goosmann} R.~W.,  {Merkulova} N.~I.,  {Shakhovskoy} N.~M.,   {Shoji} M.,  2012, \mn@doi [\apj] {10.1088/0004-637X/749/2/148}, \href {https://ui.adsabs.harvard.edu/abs/2012ApJ...749..148G} {749, 148}

\bibitem[\protect\citeauthoryear{Ghisellini, Haardt  \& Matt}{Ghisellini et~al.}{1994}]{Ghisellini1994}
Ghisellini G.,  Haardt F.,   Matt G.,  1994, \mn@doi [Monthly Notices of the Royal Astronomical Society] {10.1093/mnras/267.3.743}, 267, 743

\bibitem[\protect\citeauthoryear{{Gianolli} et~al.,}{{Gianolli} et~al.}{2023}]{Gianolli2023}
{Gianolli} V.~E.,  et~al., 2023, \mn@doi [\mnras] {10.1093/mnras/stad1697}, \href {https://ui.adsabs.harvard.edu/abs/2023MNRAS.523.4468G} {523, 4468}

\bibitem[\protect\citeauthoryear{Goosmann \& Gaskell}{Goosmann \& Gaskell}{2007}]{Goosmann2007}
Goosmann R.~W.,  Gaskell C.~M.,  2007, \mn@doi [A\&A] {10.1051/0004-6361:20053555}, \href {https://ui.adsabs.harvard.edu/abs/2007A&A...465..129G} {465, 129}

\bibitem[\protect\citeauthoryear{{Goosmann} \& {Matt}}{{Goosmann} \& {Matt}}{2011}]{Goosmann2011}
{Goosmann} R.~W.,  {Matt} G.,  2011, \mn@doi [\mnras] {10.1111/j.1365-2966.2011.18923.x}, \href {https://ui.adsabs.harvard.edu/abs/2011MNRAS.415.3119G} {415, 3119}

\bibitem[\protect\citeauthoryear{{Haardt}}{{Haardt}}{1993}]{Haardt1993}
{Haardt} F.,  1993, \mn@doi [\apj] {10.1086/173036}, \href {https://ui.adsabs.harvard.edu/abs/1993ApJ...413..680H} {413, 680}

\bibitem[\protect\citeauthoryear{{Haardt} \& {Maraschi}}{{Haardt} \& {Maraschi}}{1991}]{Haardt1991}
{Haardt} F.,  {Maraschi} L.,  1991, \mn@doi [\apjl] {10.1086/186171}, \href {https://ui.adsabs.harvard.edu/abs/1991ApJ...380L..51H} {380, L51}

\bibitem[\protect\citeauthoryear{{He}, {Liu}  \& {Zhang}}{{He} et~al.}{2016}]{He2016}
{He} J.-J.,  {Liu} Y.,   {Zhang} S.-N.,  2016, \mn@doi [\mnras] {10.1093/mnras/stv2583}, \href {https://ui.adsabs.harvard.edu/abs/2016MNRAS.455.3968H} {455, 3968}

\bibitem[\protect\citeauthoryear{{Hutsem{\'e}kers}, {Ag{\'\i}s Gonz{\'a}lez}, {Marin}, {Sluse}, {Ramos Almeida}  \& {Acosta Pulido}}{{Hutsem{\'e}kers} et~al.}{2019}]{Hutsemekers2019}
{Hutsem{\'e}kers} D.,  {Ag{\'\i}s Gonz{\'a}lez} B.,  {Marin} F.,  {Sluse} D.,  {Ramos Almeida} C.,   {Acosta Pulido} J.~A.,  2019, \mn@doi [\aap] {10.1051/0004-6361/201834633}, \href {https://ui.adsabs.harvard.edu/abs/2019A&A...625A..54H} {625, A54}

\bibitem[\protect\citeauthoryear{{Ingram} et~al.,}{{Ingram} et~al.}{2023}]{Ingram2023}
{Ingram} A.,  et~al., 2023, \mn@doi [arXiv e-prints] {10.48550/arXiv.2305.13028}, \href {https://ui.adsabs.harvard.edu/abs/2023arXiv230513028I} {p. arXiv:2305.13028}

\bibitem[\protect\citeauthoryear{{Jones}, {Gehrz}, {Kobulnicky}, {Molnar}  \& {Howard}}{{Jones} et~al.}{1994}]{Jones1994}
{Jones} T.~J.,  {Gehrz} R.~D.,  {Kobulnicky} H.~A.,  {Molnar} L.~A.,   {Howard} E.~M.,  1994, \mn@doi [\aj] {10.1086/117094}, \href {https://ui.adsabs.harvard.edu/abs/1994AJ....108..605J} {108, 605}

\bibitem[\protect\citeauthoryear{{Kallman} et~al.,}{{Kallman} et~al.}{2019}]{Kallman2019}
{Kallman} T.,  et~al., 2019, \mn@doi [\apj] {10.3847/1538-4357/ab09f8}, \href {https://ui.adsabs.harvard.edu/abs/2019ApJ...874...51K} {874, 51}

\bibitem[\protect\citeauthoryear{{Kartje}}{{Kartje}}{1995}]{Kartje1995}
{Kartje} J.~F.,  1995, \mn@doi [\apj] {10.1086/176329}, \href {https://ui.adsabs.harvard.edu/abs/1995ApJ...452..565K} {452, 565}

\bibitem[\protect\citeauthoryear{{Kravtsov} et~al.,}{{Kravtsov} et~al.}{2023}]{Kravtsov2023}
{Kravtsov} V.,  et~al., 2023, \mn@doi [arXiv e-prints] {10.48550/arXiv.2305.10813}, \href {https://ui.adsabs.harvard.edu/abs/2023arXiv230510813K} {p. arXiv:2305.10813}

\bibitem[\protect\citeauthoryear{{Krawczynski} \& {Beheshtipour}}{{Krawczynski} \& {Beheshtipour}}{2022}]{Krawczynski2022b}
{Krawczynski} H.,  {Beheshtipour} B.,  2022, \mn@doi [\apj] {10.3847/1538-4357/ac7725}, \href {https://ui.adsabs.harvard.edu/abs/2022ApJ...934....4K} {934, 4}

\bibitem[\protect\citeauthoryear{{Krawczynski} et~al.,}{{Krawczynski} et~al.}{2022}]{Krawczynski2022}
{Krawczynski} H.,  et~al., 2022, \mn@doi [Science] {10.1126/science.add5399}, \href {https://ui.adsabs.harvard.edu/abs/2022Sci...378..650K} {378, 650}

\bibitem[\protect\citeauthoryear{{Kushwaha}, {Jayasurya}, {Agrawal}  \& {Nandi}}{{Kushwaha} et~al.}{2023}]{Kushwaha2023}
{Kushwaha} A.,  {Jayasurya} K.~M.,  {Agrawal} V.~K.,   {Nandi} A.,  2023, \mn@doi [\mnras] {10.1093/mnrasl/slad070}, \href {https://ui.adsabs.harvard.edu/abs/2023MNRAS.524L..15K} {524, L15}

\bibitem[\protect\citeauthoryear{{Lightman} \& {Shapiro}}{{Lightman} \& {Shapiro}}{1975}]{LightmanShapiro1975}
{Lightman} A.~P.,  {Shapiro} S.~L.,  1975, \mn@doi [\apjl] {10.1086/181815}, \href {https://ui.adsabs.harvard.edu/abs/1975ApJ...198L..73L} {198, L73}

\bibitem[\protect\citeauthoryear{{Liu} \& {Li}}{{Liu} \& {Li}}{2014}]{Liu2014}
{Liu} Y.,  {Li} X.,  2014, \mn@doi [\apj] {10.1088/0004-637X/787/1/52}, \href {https://ui.adsabs.harvard.edu/abs/2014ApJ...787...52L} {787, 52}

\bibitem[\protect\citeauthoryear{{Loskutov} \& {Sobolev}}{{Loskutov} \& {Sobolev}}{1979}]{LoskutovSobolev1979}
{Loskutov} V.~M.,  {Sobolev} V.~V.,  1979, Astrofizika, \href {https://ui.adsabs.harvard.edu/abs/1979Afz....15..241L} {15, 241}

\bibitem[\protect\citeauthoryear{{Loskutov} \& {Sobolev}}{{Loskutov} \& {Sobolev}}{1981}]{LoskutovSobolev1981}
{Loskutov} V.~M.,  {Sobolev} V.~V.,  1981, Astrofizika, \href {https://ui.adsabs.harvard.edu/abs/1981Afz....17...97L} {17, 97}

\bibitem[\protect\citeauthoryear{Marin}{Marin}{2018}]{Marin2018}
Marin F.,  2018, \mn@doi [A\&A] {10.1051/0004-6361/201833225}, \href {https://ui.adsabs.harvard.edu/abs/2018A&A...615A.171M} {615, A171}

\bibitem[\protect\citeauthoryear{{Marin} \& {Dov{\v{c}}iak}}{{Marin} \& {Dov{\v{c}}iak}}{2015}]{Marin2015b}
{Marin} F.,  {Dov{\v{c}}iak} M.,  2015, \mn@doi [\aap] {10.1051/0004-6361/201424862}, \href {https://ui.adsabs.harvard.edu/abs/2015A&A...573A..60M} {573, A60}

\bibitem[\protect\citeauthoryear{{Marin}, {Goosmann}, {Dov{\v{c}}iak}, {Muleri}, {Porquet}, {Grosso}, {Karas}  \& {Matt}}{{Marin} et~al.}{2012a}]{Marin2012b}
{Marin} F.,  {Goosmann} R.~W.,  {Dov{\v{c}}iak} M.,  {Muleri} F.,  {Porquet} D.,  {Grosso} N.,  {Karas} V.,   {Matt} G.,  2012a, \mn@doi [\mnras] {10.1111/j.1745-3933.2012.01335.x}, \href {https://ui.adsabs.harvard.edu/abs/2012MNRAS.426L.101M} {426, L101}

\bibitem[\protect\citeauthoryear{Marin, Goosmann, Gaskell, Porquet  \& Dov{\v{c}}iak}{Marin et~al.}{2012b}]{Marin2012}
Marin F.,  Goosmann R.~W.,  Gaskell C.~M.,  Porquet D.,   Dov{\v{c}}iak M.,  2012b, \mn@doi [A\&A] {10.1051/0004-6361/201219751}, \href {https://ui.adsabs.harvard.edu/abs/2012A&A...548A.121M} {548, A121}

\bibitem[\protect\citeauthoryear{{Marin}, {Porquet}, {Goosmann}, {Dov{\v{c}}iak}, {Muleri}, {Grosso}  \& {Karas}}{{Marin} et~al.}{2013}]{Marin2013}
{Marin} F.,  {Porquet} D.,  {Goosmann} R.~W.,  {Dov{\v{c}}iak} M.,  {Muleri} F.,  {Grosso} N.,   {Karas} V.,  2013, \mn@doi [\mnras] {10.1093/mnras/stt1677}, \href {https://ui.adsabs.harvard.edu/abs/2013MNRAS.436.1615M} {436, 1615}

\bibitem[\protect\citeauthoryear{Marin, Goosmann  \& Gaskell}{Marin et~al.}{2015}]{Marin2015}
Marin F.,  Goosmann R.~W.,   Gaskell C.~M.,  2015, \mn@doi [A\&A] {10.1051/0004-6361/201525628}, \href {https://ui.adsabs.harvard.edu/abs/2015A&A...577A..66M} {577, A66}

\bibitem[\protect\citeauthoryear{{Marin}, {Dov{\v{c}}iak}, {Muleri}, {Kislat}  \& {Krawczynski}}{{Marin} et~al.}{2018a}]{Marin2018b}
{Marin} F.,  {Dov{\v{c}}iak} M.,  {Muleri} F.,  {Kislat} F.~F.,   {Krawczynski} H.~S.,  2018a, \mn@doi [\mnras] {10.1093/mnras/stx2382}, \href {https://ui.adsabs.harvard.edu/abs/2018MNRAS.473.1286M} {473, 1286}

\bibitem[\protect\citeauthoryear{{Marin}, {Dov{\v{c}}iak}  \& {Kammoun}}{{Marin} et~al.}{2018b}]{Marin2018c}
{Marin} F.,  {Dov{\v{c}}iak} M.,   {Kammoun} E.~S.,  2018b, \mn@doi [\mnras] {10.1093/mnras/sty1062}, \href {https://ui.adsabs.harvard.edu/abs/2018MNRAS.478..950M} {478, 950}

\bibitem[\protect\citeauthoryear{{Marin}, {Rojas Lobos}, {Hameury}  \& {Goosmann}}{{Marin} et~al.}{2018c}]{Marin2018d}
{Marin} F.,  {Rojas Lobos} P.~A.,  {Hameury} J.~M.,   {Goosmann} R.~W.,  2018c, \mn@doi [\aap] {10.1051/0004-6361/201732464}, \href {https://ui.adsabs.harvard.edu/abs/2018A&A...613A..30M} {613, A30}

\bibitem[\protect\citeauthoryear{{Marin}, {Le Cam}, {Lopez-Rodriguez}, {Kolehmainen}, {Babler}  \& {Meade}}{{Marin} et~al.}{2020}]{Marin2020}
{Marin} F.,  {Le Cam} J.,  {Lopez-Rodriguez} E.,  {Kolehmainen} M.,  {Babler} B.~L.,   {Meade} M.~R.,  2020, \mn@doi [\mnras] {10.1093/mnras/staa1533}, \href {https://ui.adsabs.harvard.edu/abs/2020MNRAS.496..215M} {496, 215}

\bibitem[\protect\citeauthoryear{Marinucci, Tamborra, Bianchi, Dovčiak, Matt, Middei  \& Tortosa}{Marinucci et~al.}{2018}]{Marinucci2018}
Marinucci A.,  Tamborra F.,  Bianchi S.,  Dovčiak M.,  Matt G.,  Middei R.,   Tortosa A.,  2018, \mn@doi [Galaxies] {10.3390/galaxies6020044}, 6

\bibitem[\protect\citeauthoryear{{Marinucci} et~al.,}{{Marinucci} et~al.}{2022}]{Marinucci2022}
{Marinucci} A.,  et~al., 2022, \mn@doi [\mnras] {10.1093/mnras/stac2634}, \href {https://ui.adsabs.harvard.edu/abs/2022MNRAS.516.5907M} {516, 5907}

\bibitem[\protect\citeauthoryear{{Mart{\'\i}}, {Paredes}  \& {Peracaula}}{{Mart{\'\i}} et~al.}{2000}]{Marti2000}
{Mart{\'\i}} J.,  {Paredes} J.~M.,   {Peracaula} M.,  2000, \mn@doi [\apj] {10.1086/317858}, \href {https://ui.adsabs.harvard.edu/abs/2000ApJ...545..939M} {545, 939}

\bibitem[\protect\citeauthoryear{{Matt}, {Guainazzi}  \& {Maiolino}}{{Matt} et~al.}{2003}]{Matt2003}
{Matt} G.,  {Guainazzi} M.,   {Maiolino} R.,  2003, \mn@doi [\mnras] {10.1046/j.1365-8711.2003.06539.x}, \href {https://ui.adsabs.harvard.edu/abs/2003MNRAS.342..422M} {342, 422}

\bibitem[\protect\citeauthoryear{{Miller-Jones}, {Blundell}, {Rupen}, {Mioduszewski}, {Duffy}  \& {Beasley}}{{Miller-Jones} et~al.}{2004}]{MillerJones2004}
{Miller-Jones} J. C.~A.,  {Blundell} K.~M.,  {Rupen} M.~P.,  {Mioduszewski} A.~J.,  {Duffy} P.,   {Beasley} A.~J.,  2004, \mn@doi [\apj] {10.1086/379706}, \href {https://ui.adsabs.harvard.edu/abs/2004ApJ...600..368M} {600, 368}

\bibitem[\protect\citeauthoryear{{Namekata}, {Umemura}  \& {Hasegawa}}{{Namekata} et~al.}{2014}]{Namekata2014}
{Namekata} D.,  {Umemura} M.,   {Hasegawa} K.,  2014, \mn@doi [\mnras] {10.1093/mnras/stu1271}, \href {https://ui.adsabs.harvard.edu/abs/2014MNRAS.443.2018N} {443, 2018}

\bibitem[\protect\citeauthoryear{{Narayan}, {Sa{\`I}{\textsection}dowski}  \& {Soria}}{{Narayan} et~al.}{2017}]{Narayan2017}
{Narayan} R.,  {Sa{\`I}{\textsection}dowski} A.,   {Soria} R.,  2017, \mn@doi [\mnras] {10.1093/mnras/stx1027}, \href {https://ui.adsabs.harvard.edu/abs/2017MNRAS.469.2997N} {469, 2997}

\bibitem[\protect\citeauthoryear{{Nenkova}, {Ivezi{\'c}}  \& {Elitzur}}{{Nenkova} et~al.}{2002}]{Nenkova2002}
{Nenkova} M.,  {Ivezi{\'c}} {\v{Z}}.,   {Elitzur} M.,  2002, \mn@doi [\apjl] {10.1086/340857}, \href {https://ui.adsabs.harvard.edu/abs/2002ApJ...570L...9N} {570, L9}

\bibitem[\protect\citeauthoryear{{Nenkova}, {Sirocky}, {Ivezi{\'c}}  \& {Elitzur}}{{Nenkova} et~al.}{2008a}]{Nenkova2008a}
{Nenkova} M.,  {Sirocky} M.~M.,  {Ivezi{\'c}} {\v{Z}}.,   {Elitzur} M.,  2008a, \mn@doi [\apj] {10.1086/590482}, \href {https://ui.adsabs.harvard.edu/abs/2008ApJ...685..147N} {685, 147}

\bibitem[\protect\citeauthoryear{{Nenkova}, {Sirocky}, {Nikutta}, {Ivezi{\'c}}  \& {Elitzur}}{{Nenkova} et~al.}{2008b}]{Nenkova2008b}
{Nenkova} M.,  {Sirocky} M.~M.,  {Nikutta} R.,  {Ivezi{\'c}} {\v{Z}}.,   {Elitzur} M.,  2008b, \mn@doi [\apj] {10.1086/590483}, \href {https://ui.adsabs.harvard.edu/abs/2008ApJ...685..160N} {685, 160}

\bibitem[\protect\citeauthoryear{{Oliva}, {Marconi}, {Cimatti}  \& {di Serego Alighieri}}{{Oliva} et~al.}{1998}]{Oliva1998}
{Oliva} E.,  {Marconi} A.,  {Cimatti} A.,   {di Serego Alighieri} S.,  1998, \mn@doi [\aap] {10.48550/arXiv.astro-ph/9711054}, \href {https://ui.adsabs.harvard.edu/abs/1998A&A...329L..21O} {329, L21}

\bibitem[\protect\citeauthoryear{{Podgorny} et~al.,}{{Podgorny} et~al.}{2023}]{Podgorny2023b}
{Podgorny} J.,  et~al., 2023, \mn@doi [arXiv e-prints] {10.48550/arXiv.2303.12034}, \href {https://ui.adsabs.harvard.edu/abs/2023arXiv230312034P} {p. arXiv:2303.12034}

\bibitem[\protect\citeauthoryear{{Poutanen}, {Veledina}  \& {Beloborodov}}{{Poutanen} et~al.}{2023}]{Poutanen2023}
{Poutanen} J.,  {Veledina} A.,   {Beloborodov} A.~M.,  2023, \mn@doi [arXiv e-prints] {10.48550/arXiv.2302.11674}, \href {https://ui.adsabs.harvard.edu/abs/2023arXiv230211674P} {p. arXiv:2302.11674}

\bibitem[\protect\citeauthoryear{{Ratheesh}, {Matt}, {Tombesi}, {Soffitta}, {Pesce-Rollins}  \& {Di Marco}}{{Ratheesh} et~al.}{2021}]{Ratheesh2021}
{Ratheesh} A.,  {Matt} G.,  {Tombesi} F.,  {Soffitta} P.,  {Pesce-Rollins} M.,   {Di Marco} A.,  2021, \mn@doi [\aap] {10.1051/0004-6361/202140701}, \href {https://ui.adsabs.harvard.edu/abs/2021A&A...655A..96R} {655, A96}

\bibitem[\protect\citeauthoryear{{Ratheesh} et~al.,}{{Ratheesh} et~al.}{2023}]{Ratheesh2023}
{Ratheesh} A.,  et~al., 2023, \mn@doi [arXiv e-prints] {10.48550/arXiv.2304.12752}, \href {https://ui.adsabs.harvard.edu/abs/2023arXiv230412752R} {p. arXiv:2304.12752}

\bibitem[\protect\citeauthoryear{{Rawat}, {Garg}  \& {M{\'e}ndez}}{{Rawat} et~al.}{2023}]{Rawat2023}
{Rawat} D.,  {Garg} A.,   {M{\'e}ndez} M.,  2023, \mn@doi [\apjl] {10.3847/2041-8213/acd77b}, \href {https://ui.adsabs.harvard.edu/abs/2023ApJ...949L..43R} {949, L43}

\bibitem[\protect\citeauthoryear{{Ricci} \& {Paltani}}{{Ricci} \& {Paltani}}{2023}]{Ricci2023}
{Ricci} C.,  {Paltani} S.,  2023, \mn@doi [\apj] {10.3847/1538-4357/acb5a6}, \href {https://ui.adsabs.harvard.edu/abs/2023ApJ...945...55R} {945, 55}

\bibitem[\protect\citeauthoryear{{Risaliti}, {Elvis}  \& {Nicastro}}{{Risaliti} et~al.}{2002}]{Risaliti2002}
{Risaliti} G.,  {Elvis} M.,   {Nicastro} F.,  2002, \mn@doi [\apj] {10.1086/324146}, \href {https://ui.adsabs.harvard.edu/abs/2002ApJ...571..234R} {571, 234}

\bibitem[\protect\citeauthoryear{{Rodriguez Cavero} et~al.,}{{Rodriguez Cavero} et~al.}{2023}]{Cavero2023}
{Rodriguez Cavero} N.,  et~al., 2023, \mn@doi [arXiv e-prints] {10.48550/arXiv.2305.10630}, \href {https://ui.adsabs.harvard.edu/abs/2023arXiv230510630R} {p. arXiv:2305.10630}

\bibitem[\protect\citeauthoryear{{Schawinski}, {Koss}, {Berney}  \& {Sartori}}{{Schawinski} et~al.}{2015}]{Schawinski2015}
{Schawinski} K.,  {Koss} M.,  {Berney} S.,   {Sartori} L.~F.,  2015, \mn@doi [\mnras] {10.1093/mnras/stv1136}, \href {https://ui.adsabs.harvard.edu/abs/2015MNRAS.451.2517S} {451, 2517}

\bibitem[\protect\citeauthoryear{Schnittman \& Krolik}{Schnittman \& Krolik}{2010}]{Schnittman2010}
Schnittman J.~D.,  Krolik J.~H.,  2010, \mn@doi [ApJ] {10.1088/0004-637X/712/2/908}, \href {https://ui.adsabs.harvard.edu/abs/2010ApJ...712..908S} {712, 908}

\bibitem[\protect\citeauthoryear{{Siebenmorgen}, {Heymann}  \& {Efstathiou}}{{Siebenmorgen} et~al.}{2015}]{Siebenmorgen2015}
{Siebenmorgen} R.,  {Heymann} F.,   {Efstathiou} A.,  2015, \mn@doi [\aap] {10.1051/0004-6361/201526034}, \href {https://ui.adsabs.harvard.edu/abs/2015A&A...583A.120S} {583, A120}

\bibitem[\protect\citeauthoryear{{Smith}, {Robinson}, {Alexander}, {Young}, {Axon}  \& {Corbett}}{{Smith} et~al.}{2004}]{Smith2004}
{Smith} J.~E.,  {Robinson} A.,  {Alexander} D.~M.,  {Young} S.,  {Axon} D.~J.,   {Corbett} E.~A.,  2004, \mn@doi [\mnras] {10.1111/j.1365-2966.2004.07610.x}, \href {https://ui.adsabs.harvard.edu/abs/2004MNRAS.350..140S} {350, 140}

\bibitem[\protect\citeauthoryear{{Stalevski}, {Gonz{\'a}lez-Gait{\'a}n}, {Savi{\'c}}, {Kishimoto}, {Mour{\~a}o}, {Lopez-Rodriguez}  \& {Asmus}}{{Stalevski} et~al.}{2023}]{Stalevski2023}
{Stalevski} M.,  {Gonz{\'a}lez-Gait{\'a}n} S.,  {Savi{\'c}} {\DJ}.,  {Kishimoto} M.,  {Mour{\~a}o} A.,  {Lopez-Rodriguez} E.,   {Asmus} D.,  2023, \mn@doi [\mnras] {10.1093/mnras/stac3753}, \href {https://ui.adsabs.harvard.edu/abs/2023MNRAS.519.3237S} {519, 3237}

\bibitem[\protect\citeauthoryear{{Stirling}, {Spencer}, {de la Force}, {Garrett}, {Fender}  \& {Ogley}}{{Stirling} et~al.}{2001}]{Stirling2001}
{Stirling} A.~M.,  {Spencer} R.~E.,  {de la Force} C.~J.,  {Garrett} M.~A.,  {Fender} R.~P.,   {Ogley} R.~N.,  2001, \mn@doi [\mnras] {10.1046/j.1365-8711.2001.04821.x}, \href {https://ui.adsabs.harvard.edu/abs/2001MNRAS.327.1273S} {327, 1273}

\bibitem[\protect\citeauthoryear{{Suganuma} et~al.,}{{Suganuma} et~al.}{2006}]{Suganuma2006}
{Suganuma} M.,  et~al., 2006, \mn@doi [\apj] {10.1086/499326}, \href {https://ui.adsabs.harvard.edu/abs/2006ApJ...639...46S} {639, 46}

\bibitem[\protect\citeauthoryear{{Tagliacozzo} et~al.,}{{Tagliacozzo} et~al.}{2023}]{Tagliacozzo2023}
{Tagliacozzo} D.,  et~al., 2023, \mn@doi [arXiv e-prints] {10.48550/arXiv.2305.10213}, \href {https://ui.adsabs.harvard.edu/abs/2023arXiv230510213T} {p. arXiv:2305.10213}

\bibitem[\protect\citeauthoryear{{Tamborra}, {Matt}, {Bianchi}  \& {Dov{\v{c}}iak}}{{Tamborra} et~al.}{2018}]{Tamborra2018}
{Tamborra} F.,  {Matt} G.,  {Bianchi} S.,   {Dov{\v{c}}iak} M.,  2018, \mn@doi [A\&A] {10.1051/0004-6361/201732023}, \href {https://ui.adsabs.harvard.edu/abs/2018A&A...619A.105T} {619, A105}

\bibitem[\protect\citeauthoryear{{Tristram} et~al.,}{{Tristram} et~al.}{2007}]{Tristram2007}
{Tristram} K.~R.~W.,  et~al., 2007, \mn@doi [\aap] {10.1051/0004-6361:20078369}, \href {https://ui.adsabs.harvard.edu/abs/2007A&A...474..837T} {474, 837}

\bibitem[\protect\citeauthoryear{{Unger}, {Lawrence}, {Wilson}, {Elvis}  \& {Wright}}{{Unger} et~al.}{1987}]{Unger1987}
{Unger} S.~W.,  {Lawrence} A.,  {Wilson} A.~S.,  {Elvis} M.,   {Wright} A.~E.,  1987, \mn@doi [\mnras] {10.1093/mnras/228.3.521}, \href {https://ui.adsabs.harvard.edu/abs/1987MNRAS.228..521U} {228, 521}

\bibitem[\protect\citeauthoryear{{Urry} \& {Padovani}}{{Urry} \& {Padovani}}{1995}]{Urry1995}
{Urry} C.~M.,  {Padovani} P.,  1995, \mn@doi [\pasp] {10.1086/133630}, \href {https://ui.adsabs.harvard.edu/abs/1995PASP..107..803U} {107, 803}

\bibitem[\protect\citeauthoryear{{Ursini}, {Matt}, {Bianchi}, {Marinucci}, {Dov{\v{c}}iak}  \& {Zhang}}{{Ursini} et~al.}{2022}]{Ursini2022}
{Ursini} F.,  {Matt} G.,  {Bianchi} S.,  {Marinucci} A.,  {Dov{\v{c}}iak} M.,   {Zhang} W.,  2022, \mn@doi [\mnras] {10.1093/mnras/stab3745}, \href {https://ui.adsabs.harvard.edu/abs/2022MNRAS.510.3674U} {510, 3674}

\bibitem[\protect\citeauthoryear{{Ursini} et~al.,}{{Ursini} et~al.}{2023}]{Ursini2023}
{Ursini} F.,  et~al., 2023, \mn@doi [\mnras] {10.1093/mnras/stac3189}, \href {https://ui.adsabs.harvard.edu/abs/2023MNRAS.519...50U} {519, 50}

\bibitem[\protect\citeauthoryear{{Vander Meulen}, {Camps}, {Stalevski}  \& {Baes}}{{Vander Meulen} et~al.}{2023}]{Meulen2023}
{Vander Meulen} B.,  {Camps} P.,  {Stalevski} M.,   {Baes} M.,  2023, \mn@doi [\aap] {10.1051/0004-6361/202245783}, \href {https://ui.adsabs.harvard.edu/abs/2023A&A...674A.123V} {674, A123}

\bibitem[\protect\citeauthoryear{{Veledina} et~al.,}{{Veledina} et~al.}{2023}]{Veledina2023}
{Veledina} A.,  et~al., 2023, \mn@doi [arXiv e-prints] {10.48550/arXiv.2303.01174}, \href {https://ui.adsabs.harvard.edu/abs/2023arXiv230301174V} {p. arXiv:2303.01174}

\bibitem[\protect\citeauthoryear{{Vilhu}, {Hakala}, {Hannikainen}, {McCollough}  \& {Koljonen}}{{Vilhu} et~al.}{2009}]{Vilhu2009}
{Vilhu} O.,  {Hakala} P.,  {Hannikainen} D.~C.,  {McCollough} M.,   {Koljonen} K.,  2009, \mn@doi [\aap] {10.1051/0004-6361/200811293}, \href {https://ui.adsabs.harvard.edu/abs/2009A&A...501..679V} {501, 679}

\bibitem[\protect\citeauthoryear{{Walker}}{{Walker}}{1966}]{Walker1966}
{Walker} M.~F.,  1966, \mn@doi [\aj] {10.1086/110027}, \href {https://ui.adsabs.harvard.edu/abs/1966AJ.....71Q.184W} {71, 184}

\bibitem[\protect\citeauthoryear{{Watanabe}, {Nagata}, {Sato}, {Nakaya}  \& {Hough}}{{Watanabe} et~al.}{2003}]{Watanabe2003}
{Watanabe} M.,  {Nagata} T.,  {Sato} S.,  {Nakaya} H.,   {Hough} J.~H.,  2003, \mn@doi [\apj] {10.1086/375514}, \href {https://ui.adsabs.harvard.edu/abs/2003ApJ...591..714W} {591, 714}

\bibitem[\protect\citeauthoryear{Weisskopf et~al.,}{Weisskopf et~al.}{2022}]{Weisskopf2022}
Weisskopf M.~C.,  et~al., 2022, \mn@doi [Journal of Astronomical Telescopes, Instruments, and Systems] {10.1117/1.JATIS.8.2.026002}, 8, 1

\bibitem[\protect\citeauthoryear{{Wielgus}, {Yan}, {Lasota}  \& {Abramowicz}}{{Wielgus} et~al.}{2016}]{Wielgus2016}
{Wielgus} M.,  {Yan} W.,  {Lasota} J.~P.,   {Abramowicz} M.~A.,  2016, \mn@doi [\aap] {10.1051/0004-6361/201527878}, \href {https://ui.adsabs.harvard.edu/abs/2016A&A...587A..38W} {587, A38}

\bibitem[\protect\citeauthoryear{{Wolf} \& {Henning}}{{Wolf} \& {Henning}}{1999}]{Wolf1999}
{Wolf} S.,  {Henning} T.,  1999, \aap, \href {https://ui.adsabs.harvard.edu/abs/1999A&A...341..675W} {341, 675}

\bibitem[\protect\citeauthoryear{Zhang et~al.,}{Zhang et~al.}{2016}]{Zhang2016}
Zhang S.~N.,  et~al., 2016, \mn@doi [Space Telescopes and Instrumentation 2016: Ultraviolet to Gamma Ray] {10.1117/12.2232034}

\bibitem[\protect\citeauthoryear{Zhang et~al.}{Zhang et~al.}{2019}]{Zhang2019}
Zhang S.-N.,  et~al., 2019, \mn@doi [Sci. China Phys. Mech. Astron.] {10.1007/s11433-018-9309-2}, 62, 29502

\makeatother
\end{thebibliography}




\appendix

\section{Additional modelling figures}\label{additional}

This section contains additional figures to the results presented in Sections \ref{equatorial} and \ref{cygx3_results}.

\begin{figure*}
	\includegraphics[width=2.\columnwidth]{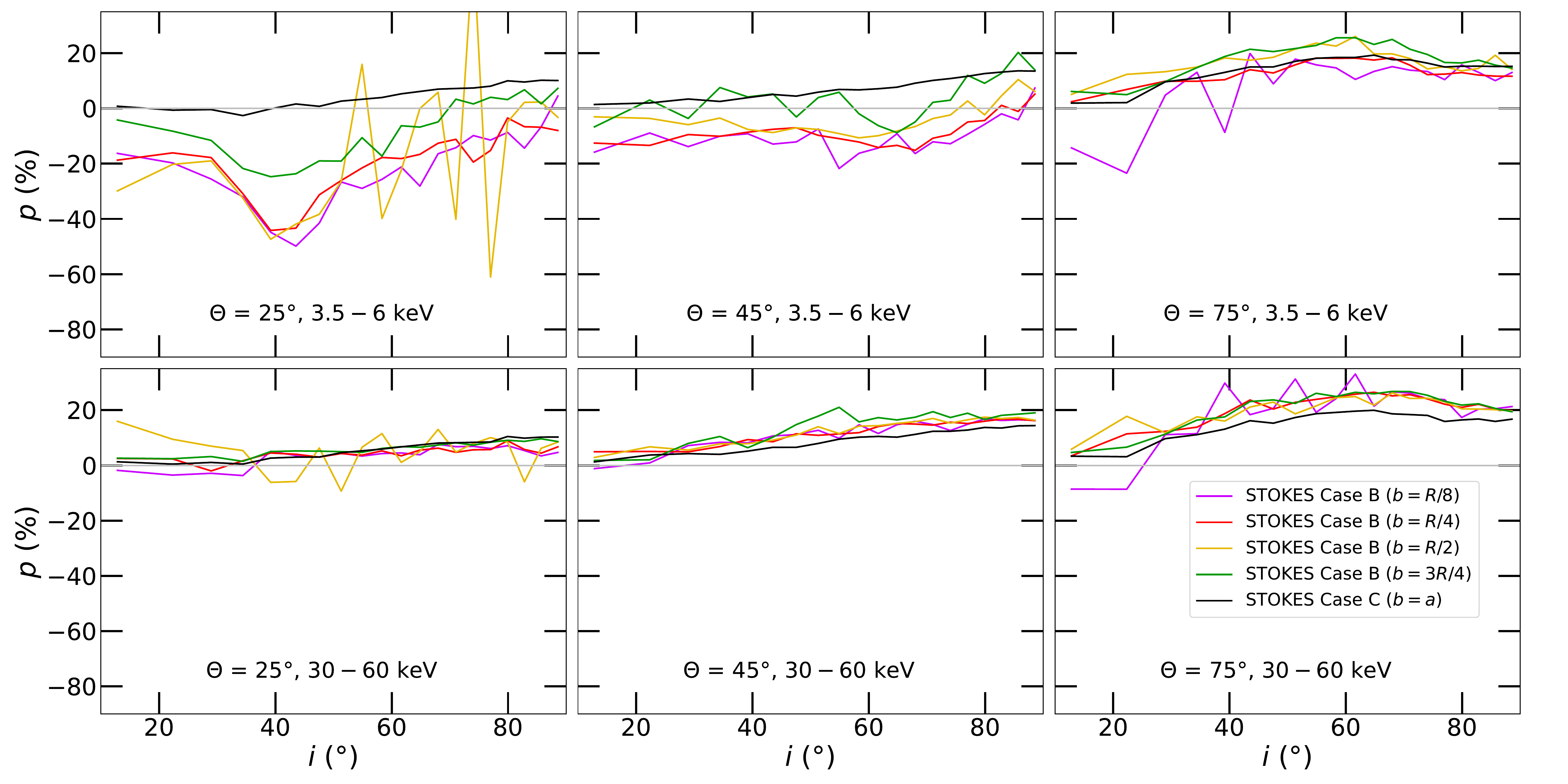}
	\caption{The same as in Figure \ref{BvsC_pmue24}, but for \textit{low} level of ionization, i.e. $\tau_\textrm{e} = 0.007$. The contribution by reflection on the furthest side and the related eccentricity effects are least prominent when free electrons are absent.}
	\label{BvsC_pmue24_low_io}
\end{figure*}
\begin{figure*}
	\includegraphics[width=2.\columnwidth]{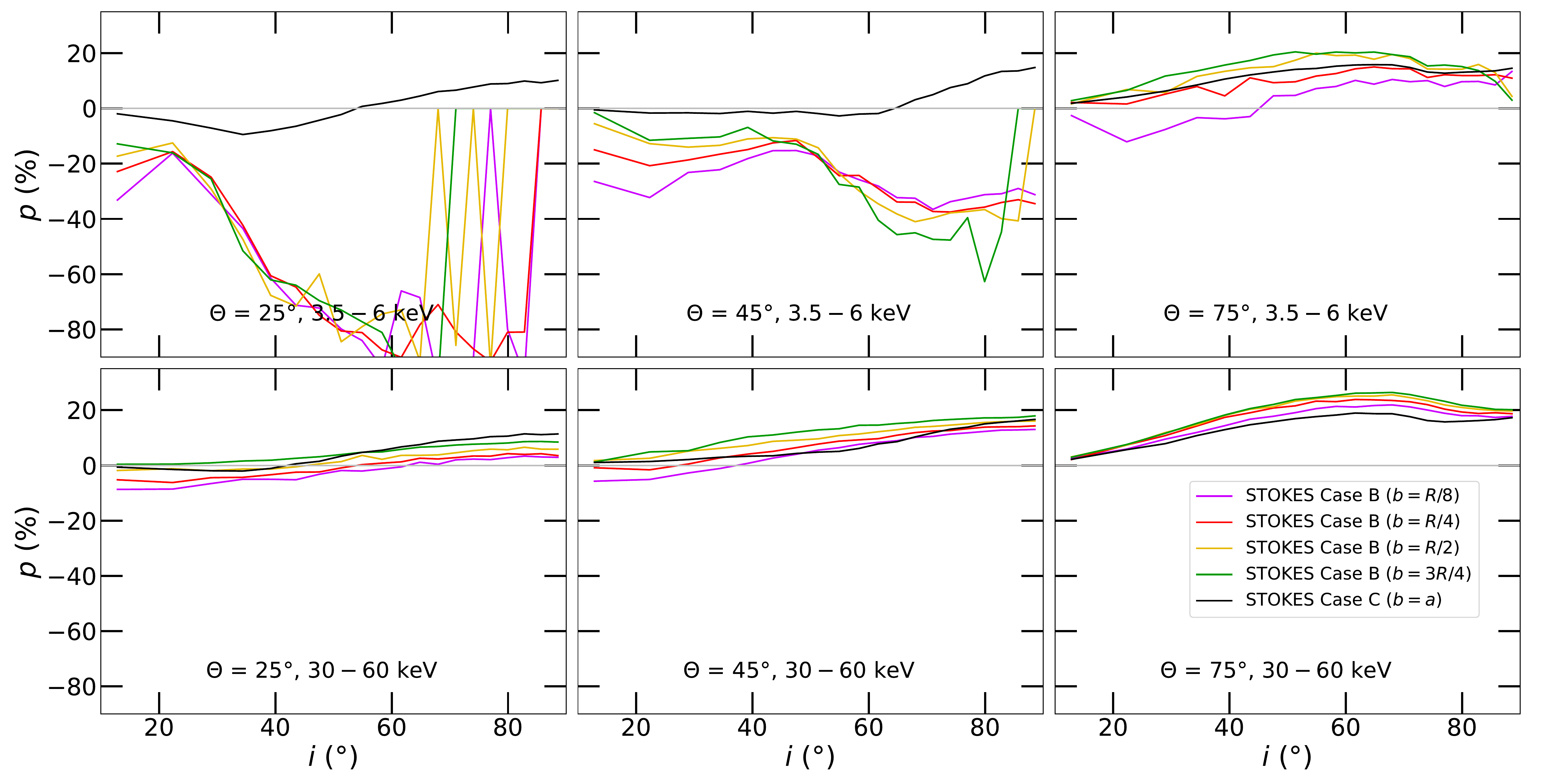}
	\caption{The same as in Figure \ref{BvsC_pmue25}, but for \textit{low} level of ionization, i.e. $\tau_\textrm{e} = 0.07$. The numerical noise is higher due to absorption, but we still observe the same eccentricity effects.}
	\label{BvsC_pmue25_low_io}
\end{figure*}
\begin{figure*}
	\includegraphics[width=2.\columnwidth]{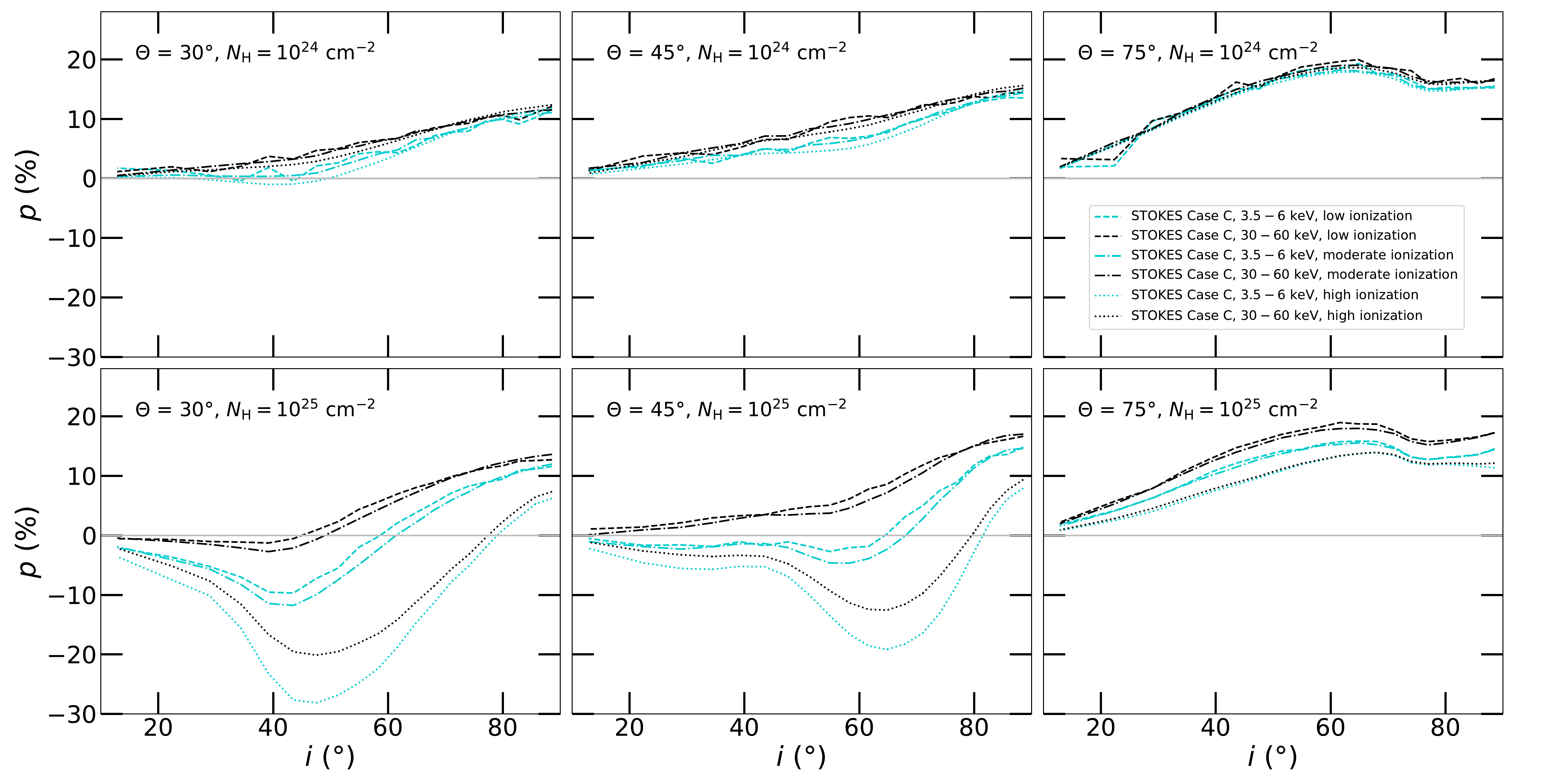}
	\caption{Case C: the polarization degree $p$ vs. inclination $i$ for $\Theta = 30\degr$ (left), $\Theta = 45\degr$ (middle), and $\Theta = 75\degr$ (right) for low, moderate and high ionization levels (i.e. dashed, dot-dashed, and dotted lines representing $\tau_\textrm{e} = 0.07, 0.7, 7$, respectively, for $N_\textrm{H} = 10^{25}  \textrm{ cm}^{-2}$ and $\tau_\textrm{e} = 0.007, 0.07, 0.7$, respectively, for $N_\textrm{H} = 10^{24}  \textrm{ cm}^{-2}$). We show the results integrated in 3.5--6 keV (blue) and 30--60 keV (black) and for $N_H = 10^{24}  \textrm{ cm}^{-2}$ (top) and $N_H = 10^{25}  \textrm{ cm}^{-2}$ (bottom). Especially for high densities, the ionization is crucial for the emergent polarization.}
	\label{CvsC_pmue}
\end{figure*}
\begin{figure*}
	\includegraphics[width=2.\columnwidth]{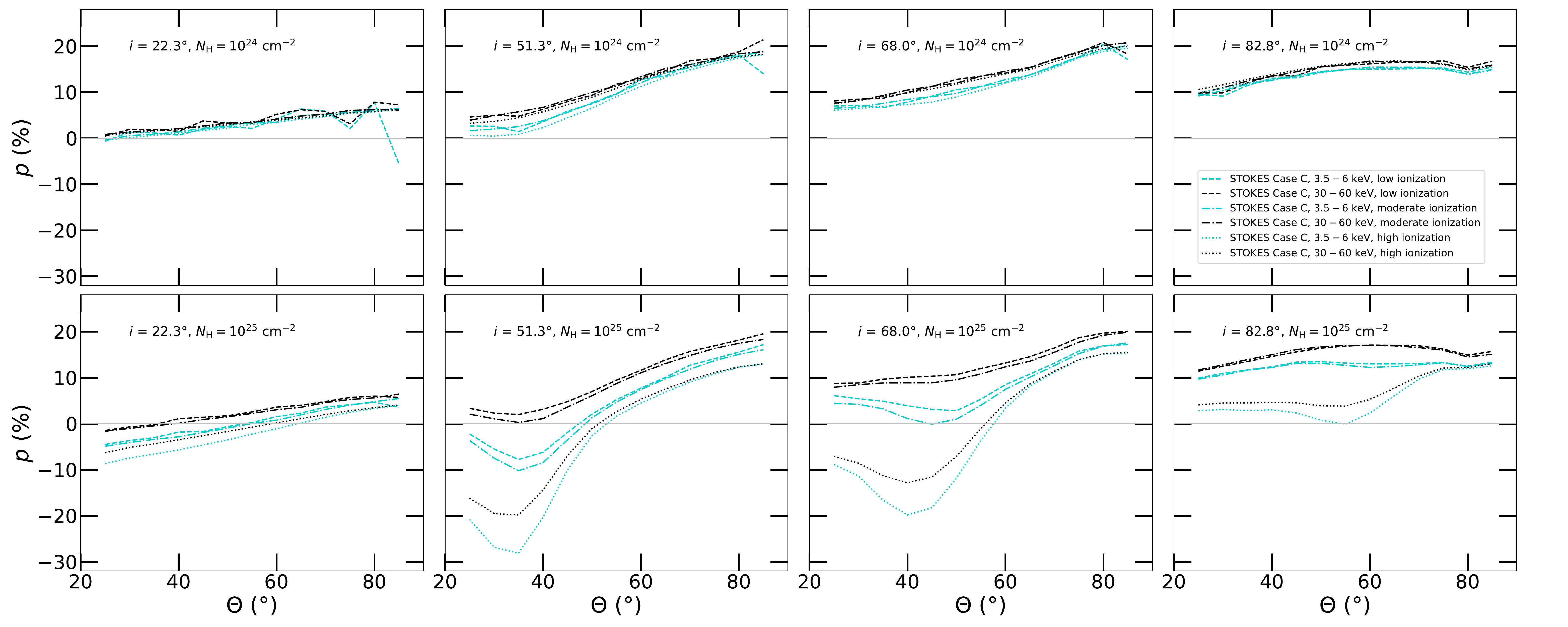}
	\caption{The same as in Figure \ref{CvsC_pmue}, but energy-averaged polarization degree, $p$, versus half-opening angle $\Theta$ is shown for $i = 22.3\degr$, $i = 51.3\degr$, $i = 68.0\degr$, and $i = 82.8\degr$ (left to right). Different view of the previous results. When absorption is high, i.e. for high densities, low half-opening angles and high inclinations, the perpendicular polarization contribution arising from inner reflection in the meridional plane is higher due to higher ionization.}
	\label{CvsC_ptheta}
\end{figure*}

\begin{figure*}
	\includegraphics[width=2.1\columnwidth]{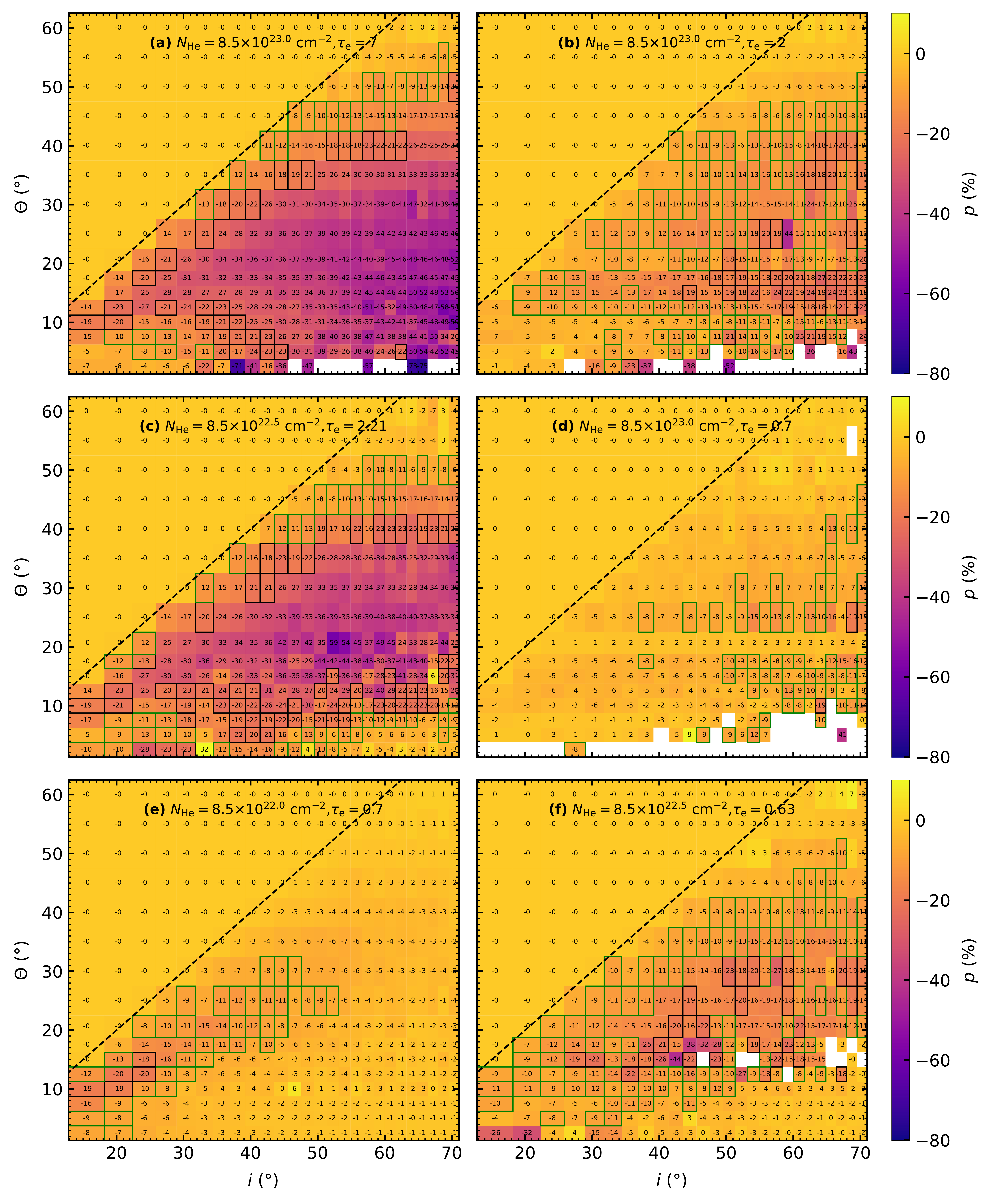}
	\caption{The same as in Figure \ref{all_torus_0.25}, but for a Case B elliptical obscurer with $b = R/8$. The two branches of solutions in the displayed configuration space remain in the same topology as on Figure \ref{all_torus_0.25}.}
	\label{all_torus_0.125}
\end{figure*}
\begin{figure*}
	\includegraphics[width=2.1\columnwidth]{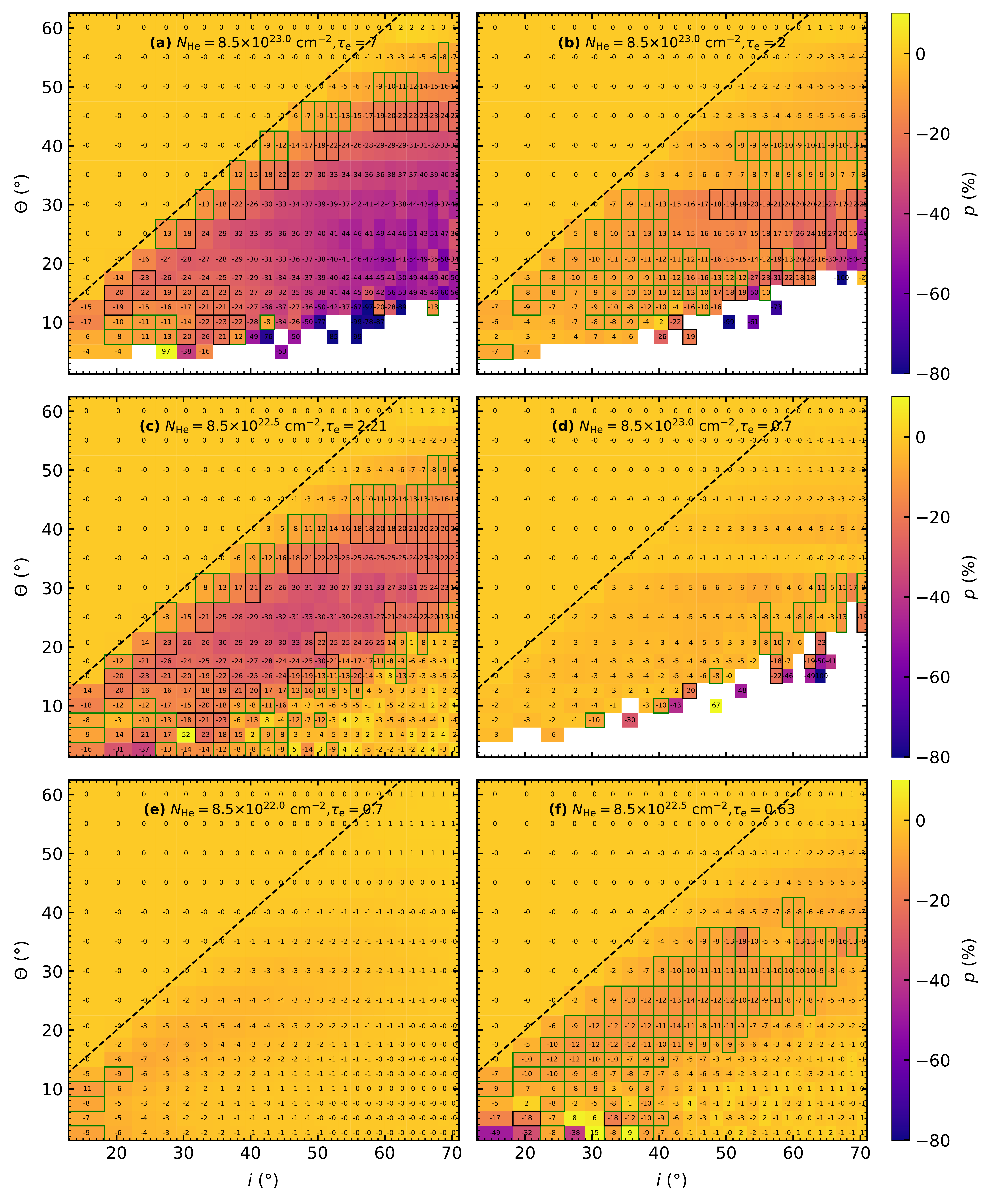}
	\caption{The same as in Figure \ref{all_torus_0.25}, but for a Case B elliptical obscurer with $b = 3R/4$. Compared to the previous cases, the line-of-sight column density is high for inclinations already just below the grazing angle (dashed line). Hence the resulting high polarization and deformation of the pattern previously observed for more prolate cases.}
	\label{all_torus_0.75}
\end{figure*}
\begin{figure*}
	\includegraphics[width=2.1\columnwidth]{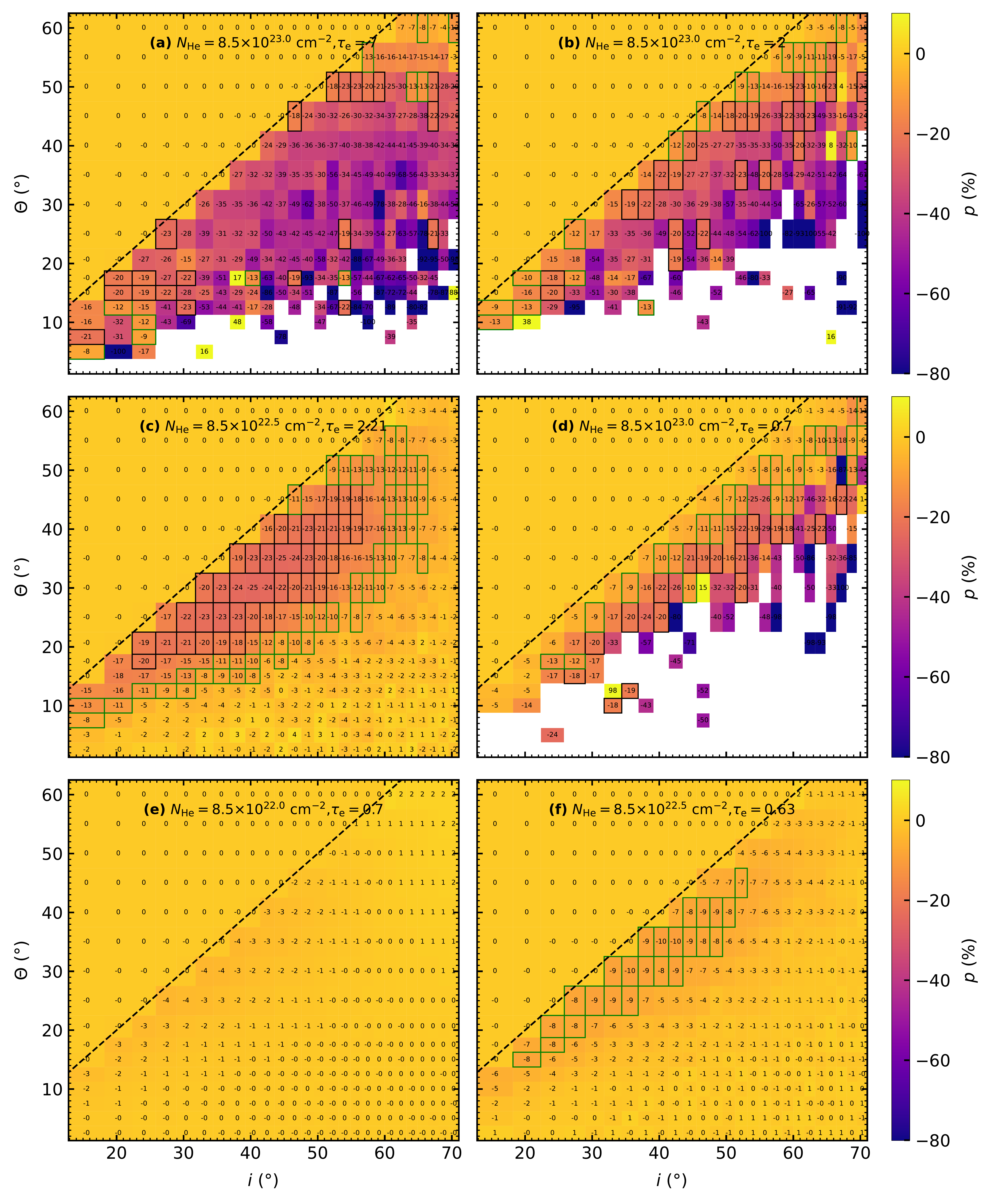}
	\caption{The same as in Figure \ref{all_torus_0.25}, but for a Case A wedge-shaped obscurer with $R_\mathrm{out} = 10$ and $r_\textrm{in} = R - b$, where $b = a$ is consistently computed through Equation \ref{opening_angle} for any circular torus, depending on inclination $i$ and half-opening angle $\Theta$. The step increase in line-of-sight column density results in high polarization for inclinations just below the grazing angle. For high inclinations and low half-opening angles the MC results suffer from low photon statistics.}
	\label{all_flared}
\end{figure*}


\bsp	
\label{lastpage}
\end{document}